\documentclass[letterpaper,twocolumn,10pt]{article}
\usepackage{usenix-2020-09}

\usepackage[ruled,vlined,linesnumbered]{algorithm2e}
\usepackage{amsmath}
\usepackage{latexsym}
\usepackage{subcaption}

\usepackage{amssymb}
\usepackage{amsthm}
\usepackage{multirow}
\usepackage{booktabs}
\usepackage{makecell}
\usepackage{caption}
\usepackage{tikz}
\usetikzlibrary{shapes.misc}
\usepackage{pgfplots}
\usepackage{pgfplotstable}
\usepgfplotslibrary{groupplots}
\usepackage[switch]{lineno}
\usepackage{enumitem}
\usepackage{authblk}

\usepackage{listings}
\usepackage{xcolor}

\lstdefinelanguage{PDDL}{
  sensitive=false,    morecomment=[l]{;}, alsoletter={:,-},   morekeywords={
    define,domain,problem,not,and,or,when,forall,exists,either,
    :domain,:requirements,:types,:objects,:constants,
    :predicates,:action,:parameters,:precondition,:effect,
    :derived,:derived-predicates,
    :fluents,:primary-effect,:side-effect,:init,:goal,
    :strips,:adl,:equality,:typing,:conditional-effects,
    :negative-preconditions,:disjunctive-preconditions,
    :existential-preconditions,:universal-preconditions,:quantified-preconditions,
    :functions,assign,increase,decrease,scale-up,scale-down,
    :metric,minimize,maximize,
    :durative-actions,:duration-inequalities,:continuous-effects,
    :durative-action,:duration,:condition,
    :constraints,preference,sometime-before
  }
}

\lstdefinelanguage{CISCO}
{
  alsoletter={:,-},
  morekeywords=[1]{router, bgp, neighbor, remote-as},
  sensitive=true,
  morecomment=[l]{!}, }

\SetCommentSty{mycommfont}

\newcommand{\name}{\textsc{Smoothie}}
\newcommand{\fd}{FastDownward}

\newcommand{\para}[1]{\vspace{0.5ex}\noindent{\bf #1}}

\newcommand{\secref}[1]{{\S\ref{#1}}}

\newcommand{\figref}[1]{Figure~\ref{#1}}

\newcommand{\tabref}[1]{Table~\ref{#1}}

\newcommand{\appref}[1]{Appendix~\ref{#1}}
\newcommand{\equref}[1]{Equation~(\ref{#1})}

\newcommand{\lstref}[1]{Listing~\ref{#1}}

\ifdefined\commentenabled
  \long\def\zp#1{{\color{red} [Peng: #1]}}
  \long\def\ratul#1{{\color{orange} [Ratul: #1]}}
  \long\def\dan#1{{\color{magenta} [Dan: #1]}}
  \long\def\todo#1{{\color{blue} [Todo: #1]}}
\else
  \long\def\zp#1{}
  \long\def\ratul#1{}
  \long\def\dan#1{}
  \long\def\todo#1{}
\fi

\usepackage{colortbl}
\usetikzlibrary{calc}

\definecolor{s2green}{RGB}{0,176,80}
\definecolor{s2red}{HTML}{F76964}
\definecolor{s2blue}{HTML}{4E83FD}
\definecolor{s2orange}{HTML}{FFA53D}
\definecolor{lightblue}{HTML}{E1EAFF}
\definecolor{lightorange}{HTML}{FEEAD2}

\newcommand{\update}[1]{#1}

\newcommand{\plan}[1]{\tikz[baseline=(ord.base)]{\node[
            rounded rectangle,
            draw=black,
            line width=0.5pt,
            fill=lightblue,
            inner xsep=3pt,
            inner ysep=2pt
        ] (ord) {\footnotesize $#1$};}}
            
\begin{document}

\title{Configuration Migration Planning through Decoupled Modeling and Searching}

\author[1,2]{Dan Wang}
\affil[1]{Xi'an Jiaotong University}
\author[1]{Peng Zhang}
\author[2]{Ratul Mahajan}
\affil[2]{University of Washington}

\maketitle

\begin{abstract}

Executing large network configuration changes, called migrations, is slow and error-prone today; operators must manually devise a safe order for updating devices that does not disrupt traffic and meets operational intents (e.g., low total time).
We develop \name, a versatile migration planning framework that, unlike existing tools, can support diverse operational intents.
Inspired by automated planning, it achieves versatility by decoupling the encoding of network safety and intents from the search for a viable plan. 
\name~uses symbolic simulation for general and scalable safety encoding, after which it can use a range of planners, including existing off-the-shelf planners and a new one we developed for optimized parallel plans.
We demonstrate \name's versatility and benchmark its performance using diverse migration scenarios on many topologies.
It finds both sequential plans and parallel plans, and the parallel ones compress up to thousands of configuration updates into no more than 3 steps.
Its versatility does not come at the expense of performance: finishing most planning tasks in less than 1 hour. 

\end{abstract}

\section{Introduction}

Modern networks are constantly evolving with \emph{large-scale} configuration change~\cite{singh2015jupiter,xu2024relational,yuan2025new,lin2025centralium}, e.g.,  
moving from iBGP full mesh to route reflection, merging two networks, and
changing the routing protocol from OSPF to BGP.
Planning such large-scale changes, termed {\em migrations}, is highly demanding~\cite{lin2025centralium}.

First, \emph{safety} is the top concern: 
since a migrations can be highly risky, e.g., affecting hundreds and even thousands of devices, 
the order of updates must ensure that the network is always in a safe state. 
For example, before turning down a link, 
operators need to ensure the traffic has been shifted to another path, in order to not disrupt the live traffic.
For another example, changing BGP configurations requires extreme carefulness to not introduce transient or persistent property violations, e.g., forwarding loops, blackholes (see \secref{sec:motivation}).

Beyond safety, operators may have \emph{diverse intent} when planning a migration.
For most of cases, they may just want to find a safe sequence as soon as possible,
and therefore a \emph{sequential plan} is enough.
Besides this, operators may require or prefer one update should take place before or after another.
In such cases, a sequential plan should not only be safe, but also satisfy some hard constraints and as many soft constraints as possible; we term such a plan as an \emph{optimal plan}.
Finally, operators of large-scale networks may desire a faster migration process, 
and therefore a \emph{parallel plan}, where a set of updates can be executed in parallel~\cite{ramanathan2023practical},
is more preferred.

Despite the dual requirements on safety and diversity,
today's configuration migration planning (CMP) is still mostly a manual effort. 
This is not only time-consuming but also error-prone.
For one example, a content service provider (\secref{sec:motivation}) reported that manual planning typically took a month or even longer.
Even after careful planning, a plan may still contain errors, which after deployment, take down a production network.
For another, 56\% of the disruptive customer-impacting incidents in Alibaba during 2016-2017 are due to configuration migration \cite{liu2018automatic}.

\begin{figure}[!t]
    \centering
    \includegraphics[width=\linewidth]{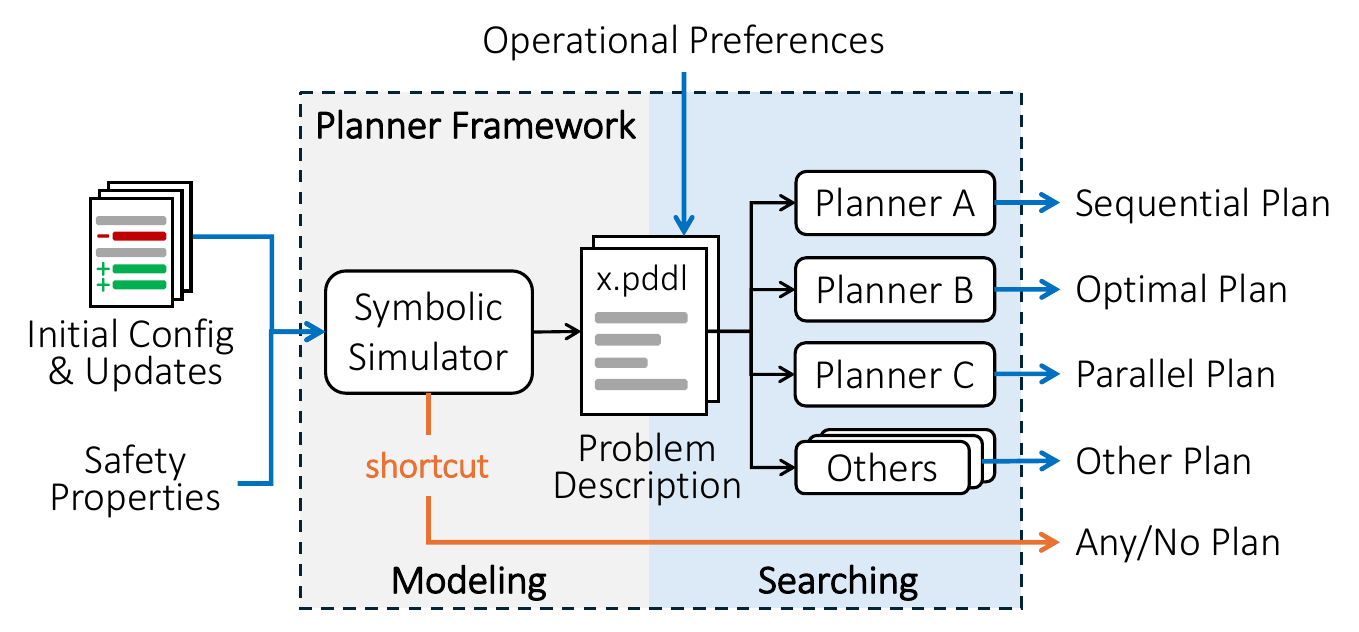}
    \caption{A decoupled framework for automated configuration migration planning.
    }
    \label{fig_planner-framework}
    \vspace{-1em}
\end{figure}

Recently, several tools have been proposed to automate the planning of configuration migration \cite{schneider2021snowcap, chen2023cursor, chen2025safemigration}.
At a high level, these tools use specific search strategies and customized network simulators to find a feasible sequence of updates.
For example, Snowcap~\cite{schneider2021snowcap} chooses one update each round, incrementally simulates the update, and validate 
the safety of the resultant state, until all updates have been applied.
Different from Snowcap which validates the safety on the fly, 
SafeMigration~\cite{chen2025safemigration} and Cursor \cite{chen2023cursor} use symbolic simulation to 
pre-identify all safe or unsafe intermediate states, and then either invoke SMT solvers or 
analyze the partial orders of updates to search for a plan.
These planners target at finding sequential plans, but fall short for finding optimal plans and parallel plans.
Also, they are faced with efficiency issues \cite{chen2025safemigration}, do not scale to large number of states \cite{schneider2021snowcap}, or are incomplete \cite{chen2023cursor}.

In this paper, we seek a principled treatment of migration planning. 
Specifically, we draw inspiration from the mature research direction of \emph{automated planning (AI planning)},
a subfield of artificial intelligence on automated synthesis of action sequences (i.e., plans).
In automated planning, a planning task \cite{vallati2016efficient,mccluskey2017embedding,shaik2023optimal}) is executed in two phases: (1) a domain-specific \emph{problem modeling} phase which models the problem in a standardized language, e.g., Planning Domain Description Language (PDDL) \cite{haslum2019pddl}, and (2) a domain-independent \emph{plan searching}, which finds a plan satisfying the model.
Such a decoupled workflow offers great flexibility, e.g, they can mix-n-match options that work best for their situation, 
add new preferences without needing to change the search strategy, or pick a different search strategy without needing to change the problem model.

Inspired by automated planning, we propose a framework for \emph{automated migration planning}.
As shown in \figref{fig_planner-framework}, the framework features \textbf{the decoupling of problem modeling and plan searching}.
In the problem modeling phase, we identify the condition for an update to be safe, 
and encode the problem in the PDDL language~\cite{aeronautiques1998pddl}. This allows us to call different backend planners for different planning tasks.

Realizing this framework is challenging.
Particularly, supporting diverse planning tasks means that we need to identify \emph{all} safe intermediate states, to ensure the searching phase can find update plans that visit only safe states; find the one with the minimum cost (for optimal plan); or identify a set of updates whose order does not matter (for parallel plan).
Clearly, using a concrete simulator such as Batfish \cite{fogel2015general}, is not realistic: 
given $n$ updates, we will have to simulate up to $2^n$ states \footnote{This paper assumes deterministic convergence. That is, given two updates $u_i, u_k$, the network will converge to the same state, either applying $u_i$ first or applying $u_k$ first. Therefore, the number of intermediates states equals the number of subsets of all updates.}.
Making the simulation incremental, as in Snowcap \cite{schneider2021snowcap}, offer limited help, as will be shown in \secref{sec:exp:simulation}.

To overcome the scalability issue, we adapt the symbolic simulation approach~\cite{ye2020accuracy,zhang2022symbolic,wang2024expresso,li2024general}, which leverages the equivalence among packet headers, link/node failures, or external routes to scale to a large search space.
Specifically, we assign each update a Boolean variable indicating whether it is applied,
and symbolically simulate the routing and forwarding behaviors of the network.
The key difference is that updates can affect the existence or value of configuration components, such as BGP sessions and IGP link costs.
We generalize SRE's way of how to initialize and transfer symbolic routes to handle symbolic configurations.

We implement a prototype of \name, and demonstrate its versatility with two planners, \fd~\cite{fast-downward-website} and ENHSP \cite{enhsp-website}, to find sequential plans. Because a suitable parallel planner does not exist, we also develop a new SMT-based planner to find parallel plans. The prototype will be available to the community upon publication.

We evaluate \name~in both real and synthetic migration scenarios. Results show that \name~generates useful plans appreciated by operators (\secref{sec:preview}), finds parallel plans that compress up to thousands of configurations updates into no more than 3 steps, while being efficient to finish most planning task in less than 1 hour (\secref{sec_exp}).

\section{Motivation}
\label{sec:motivation}

We use two concrete scenarios to motivate the need of an automatic planner for network configuration migration, 
and why the the planner should be versatile.

\subsection{Manual Migration Planning is Hard}
\label{sec_motivation-wan}

\begin{figure}[!t]
    \centering
    \includegraphics[width=\linewidth]{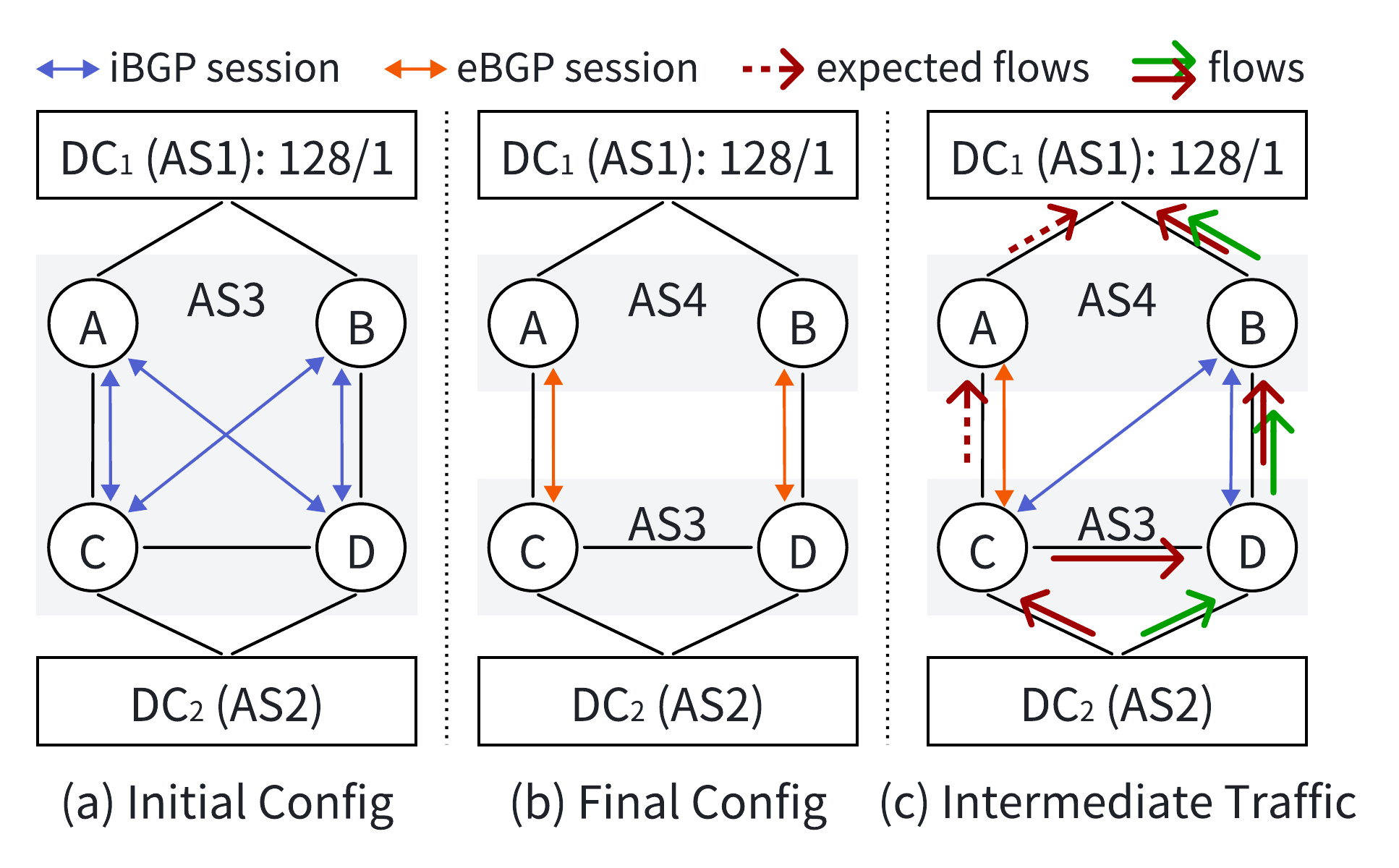}
    \caption{A simplified version of a real migration from iBGP to eBGP, in a content service provider's WAN.     
    }
    \vspace{-1em}
    \label{fig_example-wan}
    \vspace{-1em}
\end{figure}

\figref{fig_example-wan} shows a real but simplified migration scenario in a large content service provider (CSP).
The CSP has multiple datacenters across the world, interconnected with a backbone WAN. 
As shown in the figure, $A, B, C, D$ are backbone routers connecting datacenters $DC_1$ and $DC_2$.
Initially, the two backbone routers of $DC_1$ (i.e., $A, B$) and the two backbone routers of $DC_2$ (i.e., $C, D$) form pairwise iBGP sessions, as shown in \figref{fig_example-wan}(a).
One day, to improve control plane scalability, the operators decide to migrate to a new design with eBGP instead of iBGP, as shown in \figref{fig_example-wan}(b).
For this migration, the operators need to remove all iBGP sessions, 
and add two eBGP sessions.

Although it looks simple, implementing this migration without interrupting live traffic requires careful planning---it took one month for experienced operators at the CSP to make a plan! 
To illustrate this, we will use \update{$u_{xy}^+$}, \update{$u_{xy}^-$}, and \update{$u_{xy}^m$} to denote the addition, removal, and modification of the BGP session between two switches $x$ and $y$, respectively.

\emph{The Initial Plan (Incorrect).} Since $A$ (resp. $C$) and $B$ (resp. $D$) are backing up each other,
operators began with a straightforward plan \plan{u_{AC}^-, u_{AD}^-, u_{AC}^+, u_{BC}^-, u_{BD}^-, u_{BD}^+}. 
That is, first remove the two iBGP sessions and add the one eBGP session for device $A$ (i.e., \plan{u_{AC}^-, u_{AD}^-, u_{AC}^+}), and then perform similar steps for device $B$ (i.e., \plan{u_{BC}^-, u_{BD}^-, u_{BD}^+}).
However, they found that even after step \update{\update{$u_{AC}^+$}} is executed, rather than choosing the eBGP route from $A$ as expected, $C$ kept choosing the iBGP route from $B$ (due to shorter AS path).
This way, $C$ will forward traffic from $DC_2$ to $DC_1$ through the long, indirect path through $D$ and $B$ (i.e., the solid red arrows in \figref{fig_example-wan}(c)), rather than the short, direct path through $A$ (the dashed red arrows in \figref{fig_example-wan}(c)). 
Since $DC_2$ is load-balancing its traffic towards $DC_1$ to $C$ and $D$, there is inconsistency among packets sent to $C$ and $D$ (i.e., multi-path inconsistency \cite{fogel2015general}). To avoid the temporary multi-path inconsistency, the operators revise the plan as follows.

\emph{The Revised Plan (Still Incorrect).} Operators revised the above plan to perform another step \update{$u_t$} after \update{$u_{AC}^-$} and \update{$u_{AD}^-$} (i.e., \plan{u_{AC}^-, u_{AD}^-, u_t, u_{AC}^+, u_{BC}^-, u_{BD}^-, u_{BD}^+}).
That is, let $B$ increase the AS path length of routes advertised to $C$ by configuring an export policy {\tt PrependAsPath}. 
This way, after step \update{$u_{AC}^+$} is executed, $C$ will prefer eBGP routes from $A$ over the iBGP routes from $B$. 
However, this revision was ad hoc and it only solves the inconsistency after step \update{$u_{AC}^+$}; the long forwarding path still exists after step \update{$u_{AC}^-$} and before step \update{$u_{AC}^+$}. 
Plans that fundamentally avoid multi-path inconsistency are shown in \secref{sec:preview}.

\para{Summary.}
This real scenario shows that manual planning configuration migrations is difficult and error-prone; even experienced operators struggle to plan quickly and correctly.

\subsection{Diverse Migration Planning Tasks}
\label{sec_motivation-rrx2}

\begin{figure}[!t]
    \centering
    \includegraphics[width=\linewidth]{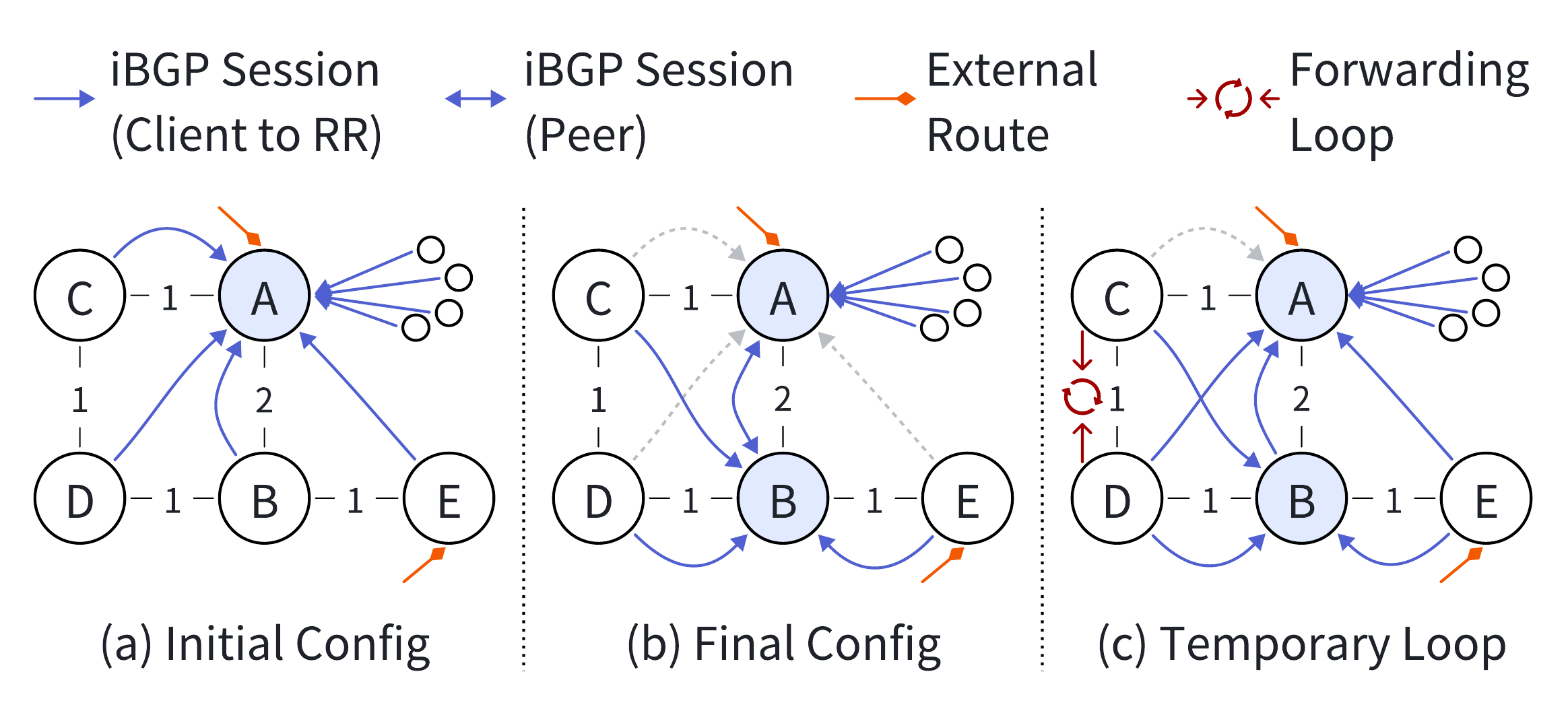}
    \caption{Example configuration migration of adding a new route reflector. Unchanged switches (i.e., persistent clients of $A$) are omitted.}
    \vspace{-1em}
    \label{fig_rrx2}
    \vspace{-1em}
\end{figure}

\figref{fig_rrx2} shows another configuration migration scenario, 
where the operators want to add a new route reflector. 
For simplicity, we only consider switches that change their route reflector.
Specifically, this migration involves three types of updates: (1) removing the client sessions from $C, D, E$ to the old route reflector $A$; (2) adding the client sessions from $C, D, E$ to the new route reflector $B$; and (3) change the session between $B$ and $A$ from Client-to-RR to Peer-to-Peer.

To ensure safety, the plan should follow the ``insert-before-remove'' principle \cite{herrero2011network} (i.e., \update{$u_{XB}^+$} before \update{$u_{XA}^-$}) for each switch to learn at least one route to the external destination.
In addition, it should enforce \update{$u_{CA}^-$} after \update{$u_{DA}^-$} to avoid forwarding loops; 
after adding the new iBGP sessions to $B$ and receiving routes from it, $C$ and $D$ still choose the route from $A$ as the best route to reach the external destination due to time-based tie-breaking for equal cost routes.
If we apply \update{$u_{CA}^-$} before \update{$u_{DA}^-$}, $C$ will learn routes only from $B$ and send the packets to $B$ through $D$, while $D$ still prefers $A$ and sends packets to it via $C$. This will result in a forwarding loop between $C$ and $D$.

There can be multiple plans satisfying safety properties.
In this example, there are a total of $(8! / 4 / 2) = 7!$ safe plans.
Instead of picking a random one from all safe plans, operators often have operational preferences over them.

For one case, they may want to apply updates switch by switch,
and therefore prefer a \emph{sequential plan} such as \plan{u_{DB}^+, u_{DA}^-, u_{CB}^+, u_{CA}^-, u_{EB}^+, u_{EA}^-}.
Moreover, operators may have some soft constraints that one update should better take place before another, 
and want to find a plan that is optimal (satisfying as many constraints as possible), 
or good enough (satisfying $\ge 3$ soft constraints). 

In another case, operators may want the migration process to be finished in shorter time, and therefore
prefer a \emph{parallel plan}, e.g., \plan{[u_{CB}^+, u_{DB}^+, u_{EB}^+], [u_{DA}^-, u_{EA}^-], [u_{CA}^-]}.
Besides being faster, a parallel plan is more flexible: they do not force a total order on independent updates, 
so that operations can flexibly pick the order of them.

\para{Summary.} On top of safety, operators can have various operational intents, and an automated planer should be versatile enough to support those diverse needs. We summarize three different planning intents:
\begin{itemize}[itemsep=1pt]
    \item \emph{Sequential Planning:} find a safe sequential migration plan as soon as possible\footnote{This is called ``agile planning'' in AI planning.};
    \item \emph{Optimal Planning:} find a safe sequential migration plan that satisfies as much operational preference as possible;
    \item \emph{Parallel Planning:} find a parallel plan that minimize the number of plan steps, thus enabling fast migration, which is a must for large-scale network.
\end{itemize}

\subsection{Limitations of Existing Work}

Existing network migration planners excel at finding any safe sequential plans, but fall short for diverse planning tasks,
due to their close bounding of specific search algorithms.
Snowcap \cite{schneider2021snowcap} offers a flexible preference definition language to enable the finding of optimal plans.
However, Snowcap adopts a counter-example guided DFS search strategy, which works well for sequential plans but does not scale well when finding optimal plans.
The reason is finding optimal plans (as well as parallel plans) require exploring a huge number of states, 
but the DFS search strategy has a complexity proportional to the number of visited states (see \secref{sec:exp:simulation}).
As reported in its paper, the number of states explored to find a local-optimal plan is quadric to the number of updates.
Cursor \cite{chen2023cursor} and SafeMigration \cite{chen2025safemigration} use dedicated search algorithms to find sequential plans, 
therefore are not amenable to efficiently support optimal and (or) parallel planning.

The research area of automated planning acknowledges the above issues.
Different planning tasks require different search strategies.
For sequential plans, the A* algorithm can find optimal plans \cite{hart1968formal,helmert2009landmarks}, while weighted A* can offer better scalability if non-optimal but ``good enough'' plans are acceptable \cite{hansen2007anytime,richter2010lama}.
Further, for parallel planning, satisfiability (SAT) or satisfiability modulo theory (SMT) based methods are a prominent approach, 
as they compactly encode concurrent action choices and avoid explicit enumeration of parallel action subsets~\cite{kautz1992planning,kautz1996pushing,rintanen2012planning}.
Therefore, we should instead develop a framework that can be easily customized for different situations with different requirements.

\subsection{\name~Preview}
\label{sec:preview}

As shown in \figref{fig_planner-framework}, \name~is a configuration migration planning framework which
decouples the plan searching and the problem modeling.
Specifically, \name~firstly symbolically simulates the control plane to obtain a migration problem, specified with a standard language (i.e., PDDL), 
such that the problem be solved by any off-the-shelf planners supporting that description language.
We tested \name~on the iBGP to eBGP migration shown in \secref{sec_motivation-wan}.
We got a sequential plan
\plan{u_{BC}^-, u_{AC}^-, u_{AD}^-, u_{AC}^+, u_{BD}^-, u_{BD}^+};
and a parallel plan 
\plan{[u_{AD}^-, u_{BC}^-], u_{AC}^-, u_{AC}^+, u_{BD}^-, u_{BD}^+}

Both plans are safe. 
\name~avoids the multi-path inconsistency by removing \update{$u_{AD}^-$} (resp. \update{$u_{BC}^-$}) before adding \update{$u_{AC}^+$} (resp. \update{$u_{BD}^+$}).
Such a principle is more obvious in the parallel plan, as it removes \update{$u_{AD}^-$} and \update{$u_{BC}^-$}, which are the root cause for longer forwarding paths, at the very beginning. 

When we showed \name's plans to the operators, they were surprised how we found them automatically.
They also pointed out that our plan would reduce the number of equivalent paths, making $DC_2$ aware of the route changes during the reconfiguration.
However, they still valued our plan much since the number of paths is not as important as the multi-path consistency property.
In fact, even after adding a preference to avoid path reduction, \name~still returned the same plans
, since the added preference is conflicting with the multi-path consistency property.

\section{Automated Planning for Configuration Migration}
\label{sec_preliminaries}

\subsection{Preliminaries to Automated Planning}
\label{sec_ai-planning}

Automated planning aims to find a valid sequence of actions, called a \emph{plan}, that transitions a system from an \emph{initial state} to a set of \emph{goal states}.
In this paper, we restrict ourselves to the classical planning problems where the state space are discrete and finite, actions are deterministic, and the system changes only through the execution of actions.

\para{Problem definition (Adapted from \cite{rintanen2012planning}).} 
\emph{An automated planning problem} is defined as a 4-tuple $(V, s_I, S_G, A)$, where $V$ is a set of Boolean \emph{state variables} and each state $s$ is a valuation of $V$, i.e., an assignment $s: V \rightarrow \{0, 1\}$; 
$s_I$ and $S_G$ are the \emph{initial state} and the set of \emph{goal states}, respectively; 
and $A$ is the set of \emph{actions} which define how a system transitions.
Each action $a \in A$ is a pair $\langle p, e\rangle$, where $p$ and $e$ are predicates over $V$, 
called \emph{precondition} and \emph{effect} of the action, respectively.
That is, an action $a = \langle p, e \rangle$ is executable at a state $s$ if $s$ satisfies $p$, i.e., $s \models p$, 
and there is 
a unique successor state $s' = exec_{\langle p, e \rangle}(s)$, where $s' \models e$ and $s'(v) = s(v)$ for all $v$ not occurring in $e$. 
This means that the effects are true in the successor state, and all state variables not affected by the action retain their values.

\para{The Language.} 
Planning Domain Definition Language (PDDL \cite{aeronautiques1998pddl}) is a widely adopted language for describing automated planning problems.
PDDL divides the model of the planning problem into two major parts:
a \emph{domain description} that defines common elements shared by all problems in a domain,
and a \emph{problem description} that defines the specific planning problem.
\lstref{lst_pddl-example} shows an example PDDL snippet, 
where the domain block (Lines 1--8) defines \emph{predicates} (symbolic, boolean-valued functions that describe properties of objects), \emph{actions}, etc.,
and the problem block (Lines 10--14) defines \emph{objects}, the initial state, and the goal states.

\para{Planners.} 
Over decades, many automated planners have emerged.
Most planners \cite{edelkamp2001mips,helmert2006fast,torralba2014symba,speck2020symk,gerevini2002lpg} are designed to find a sequential plan.
Some of them \cite{helmert2006fast} excel in finding any valid plan as quickly as possible.
Others \cite{speck2020symk,gerevini2002lpg} are good at finding a cost-optimal plan, i.e., a plan with the minimal cost, or a non-optimal but ``good enough'' plan (e.g., cost $\le 10$).
A few planners are designed to find parallel plans, but most of them conclude a set of actions can be parallelized as long as at least one total order of them is valid (known as $\exists$-step parallelism), instead of requiring any order of them to be valid (known as $\forall$-step parallelism).

\subsection{Automated CMP}
\label{sec_config-migration-planning}

\begin{lstlisting}[
  float=,
  basicstyle=\ttfamily\footnotesize,
  caption={An example PDDL file for configuration migration. Each \texttt{ui} in Line 12 represents a configuration update; the only action is to apply a given update (i.e., \texttt{apply-update} in Line 5); and \textcolor{red}{\texttt{safe}} in Line 7 is a placeholder of the formula representing all safe intermediate states where some updates are applied (i.e., \texttt{(applied ui)}) while others are not (i.e., \texttt{(\textbf{not} (applied ui))}).}, 
  label={lst_pddl-example},
  language=PDDL,
  escapechar=!,
  frame=TB,
  belowcaptionskip=-2em,
]
(define (domain config-migration)
  (:requirements :strips :typing)
  (:types bit)
  (:predicates (applied ?u - bit))
  (:action apply-update
    :parameters (?u - bit)
    :precondition (and (not (applied ?u)) !\textcolor{red}{\texttt{safe}}!)
    :effect (applied ?u)))

(define (problem instance)
  (:domain config-migration)
  (:objects u1 .. un)
  (:init (not (applied u1)) .. (not (applied un)))
  (:goal (and (applied u1) .. (applied un))))
\end{lstlisting}

We find that the configuration migration planning can be formulated as an automated planning problem, 
and such a formulation bring a lot of benefits.

\para{Problem Definition.}
A configuration migration problem with $n$ updates $\{u_1, \dots, u_n\}$ can be formulated as an automated planning problem $(V, s_I, S_G, A)$. 
\begin{itemize}[itemsep=0.5pt]
    \item $V = \{b_1, \dots, b_n\}$ is a set of Boolean variables, where $b_i$ indicates whether update $u_i$ has been applied or not, corresponding to Line 4 in \lstref{lst_pddl-example}. 
    \item $s_I = \bigwedge_{1 \le i \le n} \neg b_i$ is the initial state where no updates are applied, corresponding to Line 13 in \lstref{lst_pddl-example}.
    \item $S_G = \{ \bigwedge_{1 \le i \le n} b_i \}$ is the set of a single goal state where all updates have been applied, corresponding to Line 14 in \lstref{lst_pddl-example}.
    \item $A$ is a set of actions, where an action $\langle p_i, e_i \rangle \in A$ represents the application of update $u_i$: 
    $p_i = \neg b_i \wedge \texttt{safe}$ (the current state is safe and $u_i$ has not been applied yet; $e_i = b_i$ ($u_i$ is applied after the action. The corresponding semantics are encoded in Lines 5--8 in \lstref{lst_pddl-example}.
\end{itemize}

\para{Benefits.}
By modeling the migration planning problem as an automated planning problem with PDDL, 
we decouple the process of finding a migration plan into two phases: (1) the \emph{domain-specific} problem modeling phase, 
which identifies all safety constraints regarding network semantics with a verifier,
and (2) the \emph{domain-independent} plan searching phase, 
which enables easy plugin of existing AI planners that are optimized for diverse planning objectives (i.e., sequential, optimal, and parallel planning).
Moreover, for optimal planning, since PDDL provides a rich set of features, it is also easy to support more kinds of soft constraints.
Such as using ``temporal constraints'' to specify application durations of updates, and call dedicated planners \cite{coles2010forward,valentini2020temporal,bit2023experimenting,gerevini2002lpg} to make plans.

\subsection{Challenges and Our Approach}
\label{sec_challenges-and-approach}

\para{Challenge 1: How to identify all safe states in a large space.}
The specification of \texttt{safe} field in \lstref{lst_pddl-example} (Line 7) is critical.
To meet the need of diverse planning tasks,
the \texttt{safe} field should cover \emph{all} safe states.
Specifically, covering all safe states guarantees the completeness of finding a safe plan if one indeed exists, and sufficient information to find optimal plans and shortest parallel plans.
Using concrete simulation (e.g., Batfish \cite{fogel2015general}) is clearly unscalable, when the migration consists of $n$ updates, we have to run $2^n$ rounds of simulation.
Making the simulation incremental (as used by Snowcap \cite{schneider2021snowcap}) may speed up a single round of simulation, but does not fundamentally solve the problem, since the total time is proportional to the number of simulations (see \secref{sec:exp:simulation}).

\emph{\textbf{Our Approach: Symbolic Simulation.}}
Fortunately, there exists equivalences among all possible intermediate states.
For example, when a network migrates from iBGP full mesh to route reflection \cite{rfc-route-reflection}, most iBGP peer sessions in the full mesh carry no route exchange, and the states before and after their removals are equivalent.
We can exploit such equivalence for scalable identification of safe states using symbolic simulation \cite{zhang2022symbolic,wang2024expresso,li2024general}.
Specifically, we can symbolize the updates as symbolic configuration components, 
and walk through all states within a single symbolic simulation.
Symbolic simulation also enable a \textbf{shortcut}: 
if it verifies that \emph{all} or \emph{no} states satisfy the safety requirements,
we can quickly identify that \emph{all} or \emph{no} plans are safe (see \secref{sec:pddl:safe}).

\para{Challenge 2: How to find $\forall$-step parallel plans.} 
We find none of existing planners~\cite{kautz1992planning,rintanen2006planning,rintanen2012planning,rintanen2014madagascar,cashmore2020planning,holler2022encoding}
are suitable for finding parallel configuration migration plans, 
where a subset of updates can be executed in arbitrary order.
Some planners support $\exists$-step semantics for parallel plans, i.e., they conclude a set of actions can be parallelized when there \emph{exists} one safe order for these actions. 
Taking the migration in \figref{fig_rrx2} for example, they will parallelize $u_{CA}^-, u_{DA}^-, u_{EA}^-$ since there exists a safe order (e.g., $[u_{DA}^-, u_{CA}^-, u_{EA}^-]$), 
even though other orderings (e.g., $[u_{CA}^-, u_{DA}^-, u_{EA}^-]$) are unsafe.
Such $\exists$-step parallel planners do not meet our needs since we require \emph{all} orders of updates in the subset should be safe, 
i.e., $\forall$-step parallel plans.
Other planners can find $\forall$-step parallel plans; but prior to plan searching, they identify mutually exclusive actions if their preconditions and effects contain overlapping objects \cite{blum1997fast}.
Such a pure syntactical analysis will falsely conclude that all updates cannot be parallelized, 
while actually they are semantically independent and can be parallelized \cite{abdulaziz2023formally}.

\emph{\textbf{Our Approach: $\forall$-step Parallel Planner Identifying Semantical Independency.}}
Instead of using pure syntactic pre-processing, we semantically encode the $\forall$-step parallelism with SMT constraints.
In particular, we use the bit-vector logic and represent states, e.g., 0000 for the initial state, and 
0110 represents the 2nd and 3rd updates have been applied.
If all intermediate states, e.g., 0100 and 0010, are safe, then we conclude the 2nd and 3rd updates can be parallel.
The above semantics can be easily encoded with bit-vector operations (i.e., \texttt{bvand}). 
Details are provided in \secref{sec_parallel_planning}.

\section{Symbolic Simulation}
\label{sec_simulation}

\begin{figure*}
    \centering
    \includegraphics[width=\linewidth]{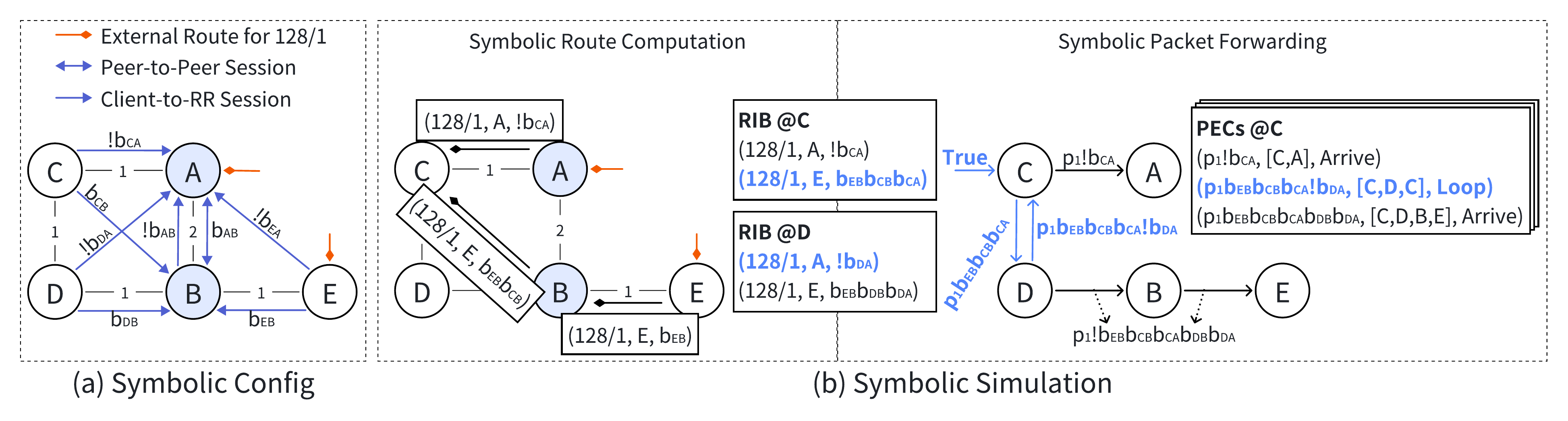}
    \vspace{-2em}
    \caption{Symbolic simulation of \name.}
    \label{fig_symbolic-simulation}
    \vspace{-1em}
\end{figure*}

This section introduces the symbolic simulation of \name. 
The overall process is similar to SRE \cite{zhang2022symbolic},
while we generalize its way to initialize and transfer symbolic routes.

\subsection{Configuration Symbolization}

\name~first computes the differences (diff) of the initial and final configurations, to get a set of updates $V = \{ u_1, \dots, u_n \}$.
Each update $u_i$ is either $\langle +, c_i \rangle$, $\langle -, c_i \rangle$, or $\langle c_i, c_i' \rangle$, indicating configuration component  $c_i$ is inserted (+), removed (-), or modified to $c_i'$, from the initial configuration to the final configuration. 
For example, the following snippet shows the configuration diff of $C$ in \figref{fig_rrx2}, and 
the corresponding two updates: (1) $u_{CA} = \langle -, \textit{iBGP}_{\textit{CA}} \rangle$ for removing the session between $C$ and $A$;
and (2) $u_{CB} = \langle +, \textit{iBGP}_{\textit{CB}} \rangle$ for inserting the session between $C$ and $B$. 

Then, \name~makes the updates and the corresponding configuration components symbolic.
For each update $u_i$, we use a boolean variable $b_i$ to represent whether it is applied ($b_i = {\tt True}$) or not ($b_i = {\tt False}$). 
Therefore, the existence of configuration component $c_i$ is symbolic: it exists if the update is an insertion that has been applied, 
or the update is removal or modification, but the update has not been applied.
\figref{fig_symbolic-simulation}(a) shows the symbolic configuration components for the example in \figref{fig_rrx2}.
We can see that the iBGP session between $C$ and $A$ exists if 
$b_{CA}$ equals \texttt{False} (the update of removal has not been applied).

\begin{lstlisting}[language=CISCO, frame=none, escapechar=?, numbers=none]
router bgp 64550
  ?\textcolor{red}{\textbf{-}}? neighbor 192.168.1.0 remote-as 64550 
  ?\textcolor{green}{\textbf{+}}? neighbor 192.168.2.0 remote-as 64550
\end{lstlisting}

\subsection{Symbolic Route Computation}

Symbolic route computation (SRC)
propagates \emph{symbolic routes} and generates \emph{symbolic RIBs}.
For each protocol, the simulation begins with switches initializing routes declared in the configurations (i.e., {\tt initialize}). Then, switches iteratively exchange routes with their neighbors (i.e., {\tt transfer}), and select the best routes from all received routes (i.e., {\tt merge}). The simulation terminates until a fix-point is reached, i.e., the RIBs of switches do not change.

We follow SRE to define a symbolic route as a 2-tuple
$R = \langle route, cond \rangle$,
where $route$ is a concrete route for a specific protocol (e.g., OSPF or BGP) specifying the prefix, next hop, and other protocol-specific attributes (e.g., AS Path); and $cond$ encodes the condition of states where $route$ becomes the best route.
Each \emph{symbolic RIB} consists of all possible routes that can materialize (i.e., become the best route) in certain states.

\para{Initializing Symbolic Routes}
is to import all routes declared in the symbolic configuration.
Different from SRE, where symbolic routes are always initialized with $cond = \texttt{True}$, in \name, the condition of an initialized symbolic route depends on whether the declaration command is symbolic or not.
If not, we set $cond = \texttt{True}$; otherwise, we set $cond = b_u$ or $cond = \neg b_u$, depending on whether the update $u$ inserts or removes the declaration command.

\para{Transferring Symbolic Routes} 
refers to transform a symbolic route when it traverses from one switch $p$ to another switch $q$,
denoted as a function $\widehat{f_{pq}}$.
Such transformation includes both route modification and condition refinement, with the route modification determined by protocol semantics and user-defined rules in configurations.

Since configurations are symbolic, we encode the symbolic transformation $\widehat{f_{pq}}$ as a collection of conditional route modifications, i.e.,
\begin{equation}
    \widehat{f_{pq}} := \{(f_{pq}^1, c_1), \dots, (f_{pq}^m, c_m) \}
\end{equation}
where $\{ f_{pq}^i \}_{i=1}^m$ is the set of all concrete transfer functions that will appear during the migration, and each $c_i$ is the encoding of all states where $f_{pq}^i$ exists, satisfying $\bigvee_{i \in [1,m]} c_i = \texttt{True}$. Correspondingly, applying $\widehat{f_{pq}}$ on a symbolic route $\langle r, c \rangle$ is defined as: 
\begin{equation}
    \widehat{f_{pq}}(\langle r, c \rangle) := \{ \langle f_{pq}^1(r), c \wedge c_1 \rangle, \dots, \langle f_{pq}^m(r), c \wedge c_m \rangle \}
\end{equation}

This encoding generalizes SRE's link-failure model, 
in which each link either applies a fixed transfer function or drops all routes, depending on its up/down state.

\name~currently considers symbolic OSPF link cost, BGP session, and routing policy.
Specifically,
if there is an update $u = \langle w, w' \rangle$ that changes the OSPF link cost between $p, q$ from $w$ to $w'$,
we have 
\begin{equation}
    \widehat{f_{pq}} = \{ (f_{pq}^w, \neg b_u), (f_{pq}^{w'}, b_u) \},
\end{equation}
where $f_{pq}^w$ (resp. $f_{pq}^{w'}$) increases the route cost by $w$ (resp. $w'$);
if there is an update $u = \langle +, session_{pq} \rangle$ that inserts a BGP session between $p$ and $q$,
we have
\begin{equation}
    \widehat{f_{pq}} = \{ (f_{pq}, b_u), (f_{\infty}, \neg b_u) \} \equiv \{ (f_{pq}, b_u) \},
\end{equation}
where $f_{\infty}$ denies all routes. It encodes the fact that the route propagation from $p$ to $q$ happens only after the session between them has established. 
For example, in \figref{fig_symbolic-simulation}(b),
$C$ will receive a route for \texttt{128/1} from $E$ after $E$ and $C$ both become clients of $B$ (i.e., $b_{EB} b_{CB}$).

We handle the removal of BGP sessions, and the modification and insertion/removal of a route policy similarly.

\para{Merging Symbolic Routes} happens when a switch learns multiple routes for the same prefix.
This step is the same as SRE: we rank these symbolic routes according to their priorities and update their conditions by negating the conditions of all higher-priority symbolic routes.
For example, in \figref{fig_symbolic-simulation}(b), 
$C$ receives two routes from $A$ and $E$, where the route from $A$ has higher priority due to lower IGP cost.
Then, the condition of the route from $E$ is updated to $b_{EB} b_{CB} \wedge \neg (\neg b_{CA}) = b_{EB} b_{CB} b_{CA}$, meaning that only after $u_{EB}^+, u_{CB}^+, u_{CA}^-$ are applied, it will become the best route.

\subsection{Symbolic Packet Forwarding}

\para{Symbolic Packet and PEC.}
Let $2^{IP}$ and $2^V$ be the universe set of all packet headers and configuration states, respectively.
A \emph{symbolic packet} is a 3-tuple $Pkt = \langle H, L, status \rangle$,
where $H \subseteq 2^{IP} \times 2^V$ is a set of header-state pairs (called symbolic header), $L$ is the forwarding path, and the forwarding $status$ is either \textsc{Fly}, \textsc{Arrive}, \textsc{Loop}, or \textsc{Drop}.
Essentially, a symbolic packet represents a set of concrete packets, which, under specific states, will traverse nodes in $L$, and have a certain $status$ at the last hop in $L$. 
Each symbolic packet not \textsc{Fly}ing is a \emph{packet equivalence class (PEC)}.

\para{Symbolic packet forwarding (SPF)}
propagates \emph{symbolic packets} throughout the network to compute a set of PECs for each switch,
in the same way as SRE \cite{zhang2022symbolic}. 
\figref{fig_symbolic-simulation}(b) shows the process for the example in \figref{fig_rrx2}.
When the symbolic packet reaches node $C$, a loop is detected, whose condition is $b_{EB} b_{CB} b_{CA} \neg b_{DA}$.

\section{PDDL Generation}
\label{sec_pddl}

The snippet in \lstref{lst_pddl-example} is a template for all configuration migrations. 
To generate PDDL for a specific migration scenario, we only need to fill in the scenario-specific safety and preference constraints in the domain file. 
It's worth mentioning that, while we focus on showing how to encode common constraints below, users are free to manipulate the generated PDDL file. 

\subsection{Encoding Safety Constraints} 
\label{sec:pddl:safe}

\emph{Safety constraints} are a set of conditions on update states (whether each update is applied or not),
under which a network satisfies a set of properties.
We term such conditions as \emph{update conditions}.
For instance, to maintain the reachability property that all switches should reach the external prefix during the migration in \secref{sec:motivation}, plans should let the network traverse only update state satisfying the condition:
\begin{equation}
\label{eq_safe}
    (\neg b_{CA} \vee b_{CB}) \wedge
    (\neg b_{DA} \vee b_{DB}) \wedge
    \neg ( b_{EB} b_{CB} b_{CA} \neg b_{DA} ),
\end{equation}
where the first and second clauses represent states where $C$ and $D$ learn at least one route for the external prefix, respectively; and the third clause encodes the states without forwarding loop. Other switches, i.e., $A, B, E$ will always learn the external prefix, thus relative variables (e.g., $b_{AB}$) do not show up in \equref{eq_safe}.

In the following, we show how to compute update conditions for different properties.
We denote properties as ${\rm Prop}(sw, p, c)$, where $sw$ is the source switch, $p$ is the destination address, and $c$ represents constraints on route attributes or packet forwarding paths. 

\para{Routing (Control Plane) Property}. A routing property ${\rm Prop}(sw, p, c)$ specifies that $sw$ should learn a route for prefix $p$ and the route attributes should satisfy a constraint $c$.
Specifically, we denote ${\rm Prop}(sw, p, {\tt True})$ as ${\rm CpReach}(sw, p)$.
The update condition for a routing property is computed as $\bigvee_{R \in \mathcal{R}} R.cond$, where
\begin{align}
    \mathcal{R} = \{ R ~|~ &R \in sw.RIB \wedge R.route.prefix = p \nonumber \\ &\wedge R.route \models c \}.
\end{align}

\para{Forwarding (Data Plane) Property}. A forwarding property ${\rm Prop}(sw, p, c)$ specifies that a packet sent from $sw$ to destination $p$ should traverse specific paths and/or have a specific final status encoded in $c$. 
Specifically, we refer to ${\rm Prop}(sw, p, status = \textsc{Arrive})$ as ${\rm DpReach}(sw, p)$; 
and the negation of ${\rm Prop}(sw, p, status = \textsc{Loop})$
as ${\rm LoopFree}(sw, p)$. 
The update condition for a forwarding property is computed as $\bigvee_{pec \in \mathcal{E}} {\tt Cond}(pec.H \wedge p)$, where
\begin{align}
    \mathcal{E} = \{ pec ~|~ & pec \in sw.PECs \wedge pec.L \models c \nonumber \\ &\wedge pec.status \models c \}.
\end{align}
and $(pec.H \wedge p)$ is a Boolean formula that consists of both IP address variables and update variables, ${\tt Cond}()$ is to extract the condition from that formula. For example, ${\tt Cond}(p_1) = {\tt True}$, ${\tt Cond}(p_1 b_1 \neg b_2) = b_1 \neg b_2$.

\para{Multi-path Consistency} specifies that all packets from $sw$ to $p$, if forwarded through multiple paths, have the same final status. We first check 
${\rm DpReach}(sw, p)$ and ${\rm Prop}(sw, p, status \in \{ \textsc{Loop}, \textsc{Drop} \})$, 
to get the condition when the destination is reached ($cond_1$) or isolated ($cond_2$), respectively.
Then, we compute the update condition as $cond_1 \wedge \neg cond_2$.

More generally, we can also check the condition when all reachable paths have the same length, as required in the example in \secref{sec_motivation-wan}; or even the condition when all reachable paths should traverse the firewall.

Since safety constraints should always be satisfied during updates,
in PDDL we encode them into the \texttt{precondition} field of the \texttt{apply-update} action (e.g., Line 7 of \lstref{lst_pddl-example}).
This will block further expansion of a plan and force back-tracing, when an unsafe state is visited. 

It is worth mentioning that if the Boolean constraint for safe states is \texttt{True} or \texttt{False}, we can directly identify that \emph{any} or \emph{none} plan is safe, respectively, without generating PDDL and calling the planners.
Even if the Boolean constraint is not \texttt{True} or \texttt{False}, it will consists of much fewer update variables than the total number of updates when most updates have no impact on safety (see \secref{sec_exp}). 
Accordingly, we call planners to plan only the updates that impact safety.

\begin{lstlisting}[
  float=,
  basicstyle=\ttfamily\footnotesize,
  caption={Encode preference in PDDL. \texttt{safe} in Line 6 is a placeholder.}, 
  label={lst_pddl-soft},
  language=PDDL,
  escapechar=!,
  frame=TB,
  belowcaptionskip=-2em,
]
(define (domain config-migration)
  (:predicates (applied ?u - bit) (violated))

  (:action apply-update
    :parameters (?u - bit)
    :precondition (and (not (applied ?u)) safe
      (not (violated)))
    :effect (and (applied ?u)
      (when (and (= ?u ui) (applied uj)) 
        (violated))))

  (:action pay-penalty
    :parameters ()
    :precondition (violated)
    :effect (and (not (violated))
      (increase (total-cost) 100))))

(define (problem instance)
  (:metric minimize (total-cost)))
\end{lstlisting}

\subsection{Encoding Preferences}

\emph{Preference constraints} are soft constraints. Their violation does not impact migration safety, but plans that satisfy more soft constraints are preferred.
Therefore, when a soft constraint is violated, instead of blocking a plan to expand, we increase the ``cost'' of the plan.
This way, the plan with the minimal cost is the plan that satisfies the most soft constraints.

Currently, we support encoding precedence preferences, i.e., preferring plans where one update $u_i$ is applied before another update $u_j$.
Inspired by \cite{wright2018compiling,percassi2019compiling}, we use \emph{conditional-effect} to encode it in PDDL, as shown in \lstref{lst_pddl-soft} (identical lines of \lstref{lst_pddl-example} omitted for brevity).
In detail, we first define a predicate \texttt{violated} to flag preference violation (Line 2), 
and then define conditional effect (Lines 9 - 10), which encodes that the preference is violated if the current update is $u_i$, and $u_j$ is already applied.
If violated, the plan is blocked from expanding (Line 7) unless the penalty for preference violation is paid (Lines 12 - 16).
Finally, we inform the planner to find the plan with the minimal cost (Line 19).
If there are multiple such preferences, we encode their disjunction in Line 9.

\subsection{Optimizations} 

For configuration migration, a common safety requirement is to maintain global (all switches') reachability to specific destinations.
However, we found that a direct computation of global reachability conditions using \equref{eq_dp-reach} leads to a huge \texttt{safe} constraints, 
making plan searching untractable.

To reduce the size of constraints, we introduce an optimization termed \emph{safety constraint simplification}.
This optimization converts the global reachability property for a prefix $p$ into an equivalent one, i.e., control plane reachability plus data plane loop freedom.
Formally, instead of computing: 
\begin{equation}
\label{eq_dp-reach}
    \bigwedge_{n_1 \in \mathcal{N}} {\rm DpReach}(n_1)
\end{equation}
where $\mathcal{N}$ is the set of all switches, we compute:
\begin{equation}
\label{eq_cp-reach}
    (\bigwedge_{n_1 \in \mathcal{N}} {\rm CpReach}(n_1)) \wedge (\bigwedge_{n_1 \in loops} {\rm LoopFree}(n_1))
\end{equation}

If no data plane packet filtering or rewriting configuration (e.g., ACL, NAT) exists,
these two formulas are equivalent since the DP reachability of $n_1$, i.e., ${\rm DpReach}(n_1)$, depends recursively on the DP reachability of its next-hops:
\begin{align}
\label{eq_dp-nh}
    {\rm DpReach}(n_1) \leftrightarrow
    \bigvee_{(n_1, n_2) \in E} ( & {\rm CpReach}(n_1, n_2) \wedge {\rm DpReach}(n_2) \nonumber \\
    & \wedge \neg {\rm DpReach}(n_2, n_1) ),
\end{align}
where $E$ is the link set of the network, ${\rm CpReach}(n_1, n_2)$ represents $n_1$ selects $n_2$ as its next-hop, and ${\rm DpReach}(n_1, n_2)$ represents $n_2$'s data plane reachability via forwarding paths that traverses $n_1$.
Specifically, the negation of ${\rm DpReach}(n_1, n_2)$ prevents forwarding loops,
which is a detailed counterpart of the loop freedom encoding in \equref{eq_cp-reach}.

While equivalent, \equref{eq_dp-reach} usually results in more complicated formulas than \equref{eq_cp-reach}, which will impede planners like \fd. Using \equref{eq_cp-reach} directly frees \name~from minimizing the formulas.
\section{Plan Searching}
\label{sec_parallel_planning}

After obtaining the PDDL, we can invoke off-the-shelf planners based on operators' intent.
Specifically, we can use \fd~\cite{fast-downward-website} to find a sequential plan as soon as possible;
if some soft constraints are appointed, we can use planners like ENHSP~\cite{enhsp-website} to find the sequential plan that satisfies the most soft constraints.
Since no existing planners can find $\forall$-step parallel plan for our problem,
we customize our own planner based on SMT.

\para{Customized SMT-based Planner.} Suppose there are $n$ updates, our planner finds the \emph{shortest} plan by checking whether there exists a $k$-step plan for an increasing $k$ ($1 \le k \le n$).
For each $k$, we convert the problem of searching a $k$-step plan into a set of SMT constraints:
\begin{equation}
\label{eq:parallel}
    I(s_0) \wedge {\tt trans}(s_0, s_1) \wedge \cdots \wedge {\tt trans}(s_{k-1}, s_{k}) \wedge G(s_{k}),
\end{equation}
where $s_i$ represents the state after $i$ steps, $I(s_0)$ and $G(s_{k})$ constraints $s_0$ and $s_{k}$ to be the initial and goal state, respectively; ${\tt trans}(s_i, s_{i + 1})$ ($i \in [0, k-1]$) constraints the transition between states.

We represent states with bit-vectors of length $n$, with the $i$th position represents whether the $i$th update has been applied.
$s_j$ is said to be a successor of $s_i$, denoted as $s_i \prec s_j$, if all updates that have been applied in state $s_i$, have also been applied in state $s_j$, i.e.,
\begin{equation}
    s_i \prec s_j := \left( {\tt bvand}(s_i, s_j) = s_i \right) \wedge \left( s_i \neq s_j \right).
\end{equation}
Then, a state $s'$ is said to be an intermediate state of $s_i$ and $s_j$ if and only if $s_i \prec s' \prec s_j$.

Under $\forall$-step parallelism semantics, 
the updates corresponding to bits that change from $0$ in $s_i$ to $1$ in $s_{i+1}$ must be safely executable in any order.
Equivalently, every intermediate state between $s_i$ and $s_j$ must be safe:
\begin{equation}
    {\tt parallel}(s_i, s_{i+1}) := \forall_{s_i \prec s \prec s_{i+1}} {\tt safe}(s),
\end{equation}

Finally, the \texttt{trans} constraints is encoded as:
\begin{equation}
    {\tt trans}(s_i, s_j) := (s_i \prec s_j) \wedge {\tt parallel}(s_i, s_j).
\end{equation}

\section{Evaluation}
\label{sec_exp}

Our evaluation aims to answer the following questions: 
\textbf{(Q1)} Can \name~find high quality plans? 
\textbf{(Q2)} How \name~perform on different planning tasks? 
\textbf{(Q3)} How \name~compares to state-of-the-art planners?
\textbf{(Q4)} Can symbolic simulation scale?

\para{Implementation and Setup.} 
We implemented a prototype of \name~with around 25k lines of Java code. 
For the symbolic simulator of \name, we use BDD, specifically the JDD library~\cite{jdd}, to encode Boolean formulas and operations. 
For the planning backend, we use \fd~\cite{fast-downward-website}, ENHSP \cite{enhsp-website}, and 
our parallel planner (\secref{sec_parallel_planning}) with Z3 \cite{z3} to find sequential and parallel plans, respectively.
All evaluations are executed on a Linux server with two 16-core Intel Xeon CPUs @ 2.90GHz and 256G RAM.

\subsection{Migration Scenarios}
\label{sec_dataset}

We use 87 topologies from Topology Zoo \cite{knight2011internet}.
The topologies have 5--82 switches (avg. 33), and 4--96 edges (avg. 39).
For each topology, we configure 
(\textit{i}) OSPF with uniform link cost; 
(\textit{ii}) iBGP route reflection, where the switch with the maximum edge degree is the route reflector; and
(\textit{iii}) all external switches to announce the same prefix. 

We consider 5 synthetic migration scenarios.
Except \texttt{RRx2}, all of them are from Snowcap~\cite{schneider2021snowcap}.
The first three of the five scenarios are inspired by real-world migrations, 
while the last two are mainly for benchmarking. 
For all scenarios, we consider global reachability (\secref{sec_pddl}) as the safety requirement, i.e., every internal switch can reach the external prefix.

\para{(1) RRx2} 
adds a new route reflector.
We randomly partition the network into two regions that are balanced in the number of switches using METIS \cite{metis}.
For the region that does not contain the original reflector $RR_1$, we select a new reflector $RR_2$.
For each switch other than $RR_2$ in the second region, we delete its session to $RR_1$, and add a new session to $RR_2$.
Thus, there are $O(n)$ updates for a network with $n$ switches.

\para{(2) FM2RR} 
transforms from iBGP full mesh into route reflection.
For a topology with $n$ switches, there are $O(n^2)$ updates, with removing $O(n^2)$ iBGP sessions from the full mesh, and adding $O(n)$ reflector-client sessions.

\para{(3) NetAcq}
merges two networks, a newly acquired network and an existing network, into a single one.
During the migration, we adjust the OSPF cost of one network to the same value with the other, and configure the route reflectors of the two networks to set up an iBGP peer-to-peer session.
The number of updates is also proportional to the number of edges in the topology.
Only 62 topologies are eligible for this scenario since it requires one external switch for each network.

\para{(4) IGPx2}
doubles the OSPF cost for each link in the network.
There are $m$ updates for a topology with $m$ edges.

\para{(5) LPx2}
doubles the local preference for each external neighbor of the network.
If the network has $p$ external neighbors, then there will be $p$ updates.

For all following figures using ``Instance ID'' as xtics, the topologies are sorted by increasing number of updates.

\subsection{Plan Qualities}
\label{sec:exp:plan-quality}

We show qualities of the plans found by \name.

\begin{table}[]
    \footnotesize
    \centering
    \begin{tabular}{l|c|c|c|c|c}
        \toprule
        & \texttt{RRx2} & \texttt{FM2RR} & \texttt{NetAcq} & \texttt{IGPx2} & \texttt{LPx2}  \\
        \midrule
        \#total & 82 & 87 & 62 & 87 & 87 \\
        \#all-safe & 2 & 42 & 62 & 87 & 87 \\
        \bottomrule
    \end{tabular}
    \caption{
    Statistics for the five migration scenarios. \#total means the total number of instances; \#all-safe means the number of instances that symbolic simulation result indicates all intermediate states are safe.
    }
    \label{tab_overall}
    \vspace{-1em}
\end{table}

\para{Overall Statistics.}
As shown in \tabref{tab_overall}, for 42 (2) \texttt{FM2RR} (\texttt{RRx2}) instances, and all instances of the other three scenarios, 
symbolic simulation concludes that all intermediate states satisfy the global reachability, i.e., 
all plans are safe, and there is no need to call a planner. 
In the following, we show details about the instances that really need to call a planner.

\begin{figure}
    \centering
    \includegraphics[width=\linewidth]{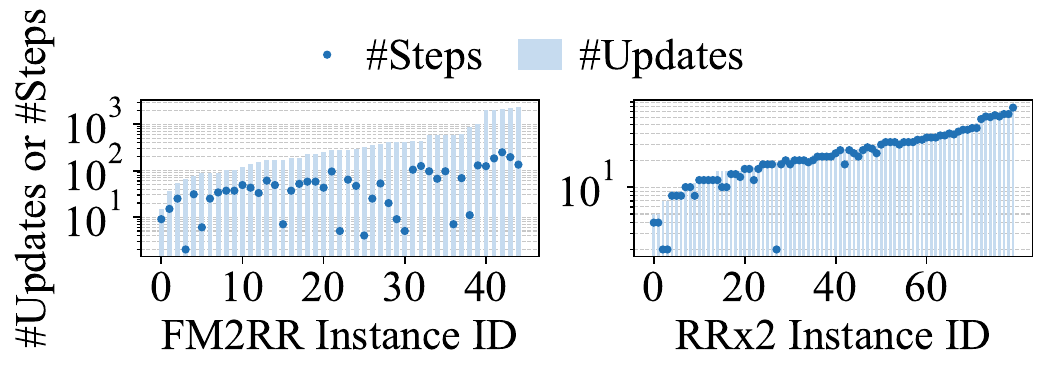}
    \vspace{-2em}
    \caption{Number of updates and sequential plan steps for \texttt{FM2RR} and \texttt{RRx2} instances.}
    \label{fig_seq-length}
    \vspace{-1em}
\end{figure}

\para{Sequential Plans.} 
\figref{fig_seq-length} shows the number of updates and sequential steps for each instance that needs planning.
For \texttt{FM2RR}, the sequential plans have up to 2 magnitudes fewer steps than updates. 
This is because most updates are removing iBGP peer sessions that have no impact on safety: the end switches of such sessions only receive iBGP routes from the border switches that peer with external neighbors, or from the route reflector; therefore, they do not exchange routes with each other upon such sessions.
On the contrary, for most \texttt{RRx2} instances, the number of steps is roughly the same as the number of updates, since most updates will impact safety.
Moreover, except for 5 timeouts, ENHSP successfully finds the optimal plan for all other instances.

\begin{figure}
    \centering
    \vspace{-1em}
    \includegraphics[width=\linewidth]{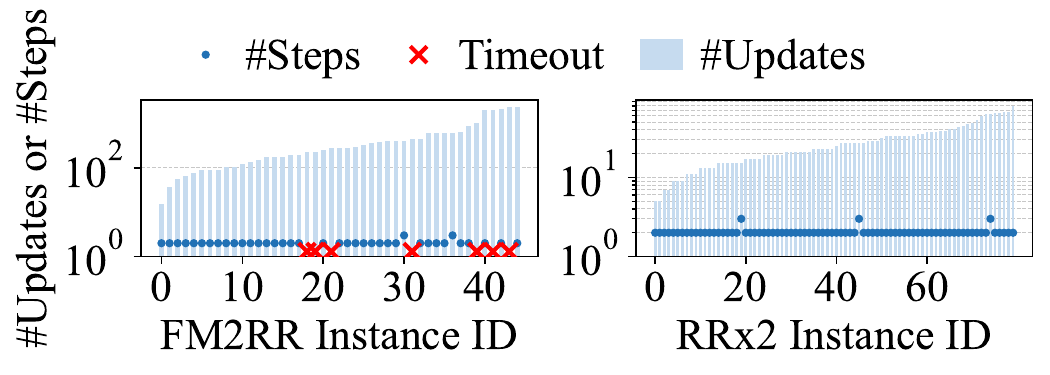}
    \vspace{-2em}
    \caption{Number of updates and parallel plan steps for \texttt{FM2RR} and \texttt{RRx2} instances.}
    \label{fig_para-length}
    \vspace{-1em}
\end{figure}

\para{Parallel Plans.}
\figref{fig_para-length} shows the number of updates and parallel plan steps for each instance.
For all non-timeout instances, the shortest parallel plan contains only $\le 3$ steps!
Compressing hundreds or thousands of updates into $O(10)$-step parallel plans enables fast migration, which is a big win for highly automated production networks \cite{ramanathan2023practical}.
However, chances are certain migrations may be too complicated for the parallel planner to find the shortest parallel plan, as indicated by the timeouts. 
When that happens, we can trade off the plan length and try again, or fall back on using the more scalable sequential planners.

\begin{figure}[!t]
    \centering
    \includegraphics[width=\linewidth]{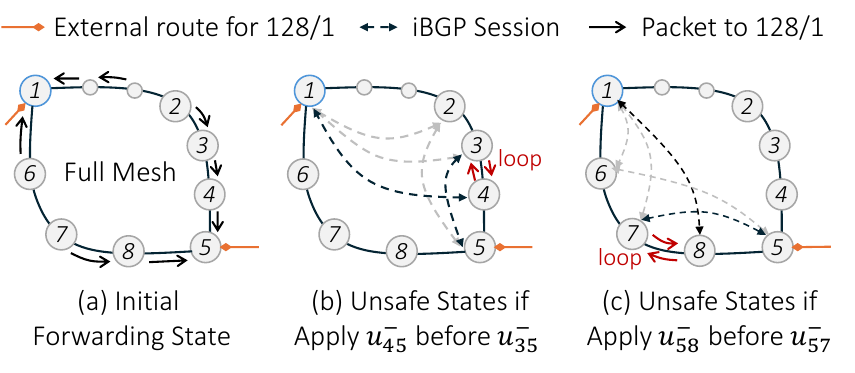}
    \vspace{-2em}
    \caption{The initial and unsafe states when applying \texttt{FM2RR} on GtsPoland, with Node 1 as the route reflector. Irrelevant nodes, iBGP sessions, and traffic are removed for simplicity.}
    \label{fig:exp:fm2rr}
    \vspace{-1em}
\end{figure}

\para{Example.}
\figref{fig:exp:fm2rr} illustrates applying \texttt{FM2RR} on GtsPoland, whose shortest parallel plan has three steps.
In the initial full mesh state, packets to the external prefix flows alone $n_2 \rightarrow n_3 \rightarrow n_4 \rightarrow n_5$.
During the migration to the route reflection state, removing the iBGP session between $n_4$ and $n_5$ (i.e., $u_{45}^-$) before removing the one between $n_3$ and $n_5$ (i.e., $u_{35}^-$) will cause forwarding loop between $n_3$ and $n_4$, as shown in \figref{fig:exp:fm2rr}(b).
Therefore, a safe plan should apply $u_{35}^-$ before $u_{45}^-$.
Similarly, $u_{25}^-$ (resp. $u_{57}^-$) should be applied before $u_{35}^-$ (resp. $u_{58}^-$);
For this instance, Snowcap outputs a sequential plan \plan{\dots, u_{25}^-, \dots, u_{35}^-, \dots, u_{57}^-, \dots, u_{45}^-, \dots, u_{58}^-, \dots}.
Even though safe, it is too concrete to reveal the fact that ``updates other than $u_{25}^-, u_{35}^-, u_{45}^-, u_{57}^-, u_{58}^-$ do no impact safety, and can be randomly ordered''.
In contrast, \name~outputs a sequential plan \plan{u_{25}^-, u_{35}^-, u_{57}^-, u_{45}^-, u_{58}^-}, 
or a parallel plan \plan{[u_{25}^-],[u_{35}^-, u_{57}^-], [u_{45}^-, u_{58}^-]}, 
indicating that the order of all other updates does not matter.

\subsection{Performance on Different Tasks}
\label{sec:exp:planners}

This section evaluates how \name~performs on different planning tasks,
when equipped with different planners.

\begin{figure}
    \centering
    \includegraphics[width=\linewidth]{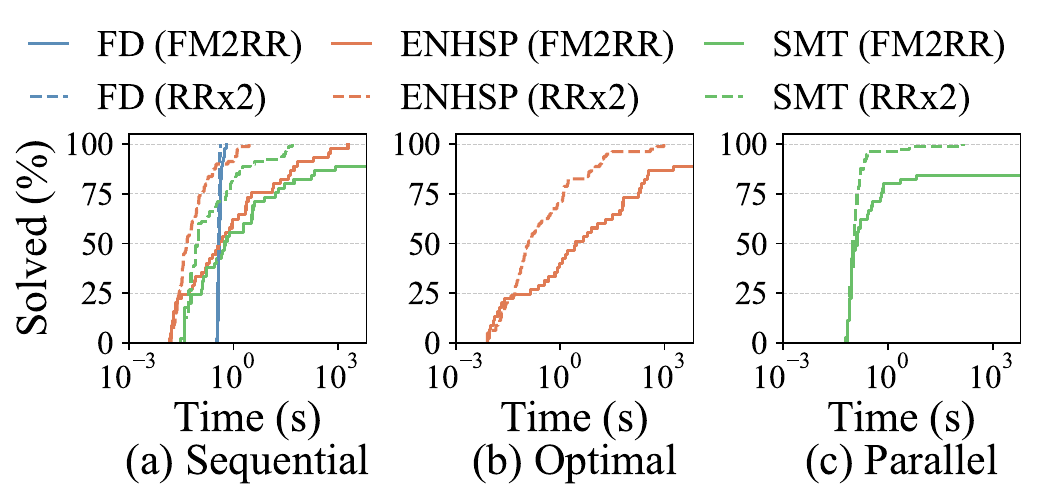}
    \vspace{-2em}
    \caption{The searching time when using different planners for three types of planning tasks.}
    \label{fig_planning-time-cdf}
    \vspace{-1em}
\end{figure}

\para{Sequential Planning.}
\figref{fig_planning-time-cdf}(a) shows the results for finding sequential plans using all three planners.
In general, \fd~ is extremely scalable, finishes in less than one second for all instances;
ENHSP is also scalable, completing within 100 seconds in most instances; 
while the SMT planner has the worst scalability, taking around 1000 seconds, or even runs out of time, on some complex instances.

\para{Optimal Planning.}
\figref{fig_planning-time-cdf}(b) shows the results for appointing a random precedence preference and uses ENHSP to find the optimal plan, which is the only planner that supports such planning tasks.
Compared to sequential planning, ENHSP is slightly slower for optimal planning; 
but it still completes the search on most instances within 100 seconds.

\para{Parallel Planning.} 
\figref{fig_planning-time-cdf}(c) shows the results for finding parallel plans using our SMT-based planner, the only planner that supports parallel planning. 
Our planner can find the shortest parallel plan in less than 1 minute, except for running out of time on 7 \texttt{FM2RR} instances. 
Interestingly, for around 45\% of all instances, our planner is more efficient to find parallel plans than sequential plans.
This is mainly because finding sequential plans requires unrolling ${\tt trans}$ in \equref{eq:parallel} $n$ times (where $n$ is the number of updates), making the resulting SMT constraint too large to be solved efficiently.

In sum, the results show that different planners excel on different planning tasks, 
no single planning strategy suites all operational requirements.
This indicates a decoupled planning framework is necessary for support diverse planning tasks.

\subsection{Comparison with SOTA}
\label{sec:exp:comparison}

\begin{figure}[!t]
    \centering
    \includegraphics[width=\linewidth]{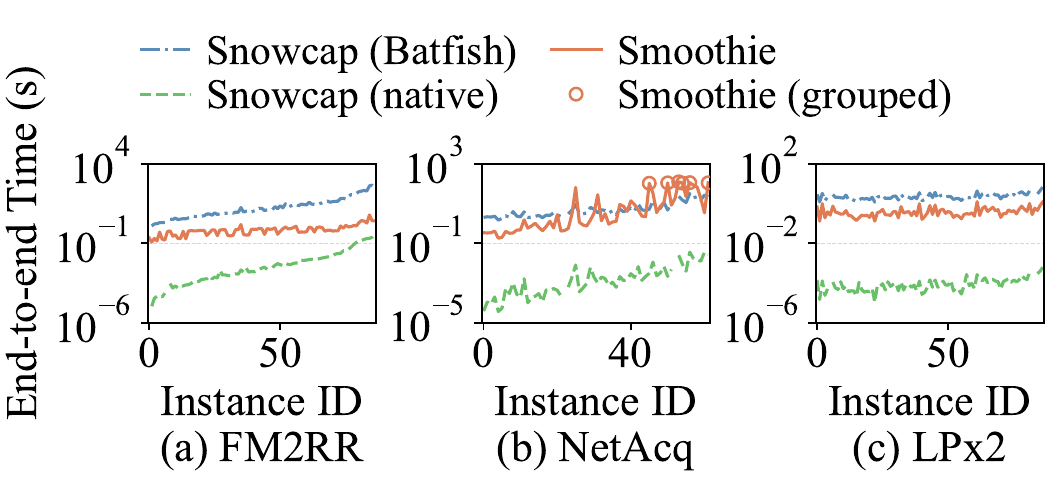}
    \vspace{-2em}
    \caption{
    End-to-end time to find any safe sequential plan using Snowcap (native), Snowcap (Batfish), \name, and \name~(grouped). 
    }
    \vspace{-2em}
    \label{fig:e2e}
\end{figure}

This section compares \name~with Snowcap~\cite{snowcap_prototype}, the state-of-the-art configuration migration planner. 
We compare their time to find any sequential plan.

As shown in \figref{fig:e2e},
Snowcap is extremely fast. It takes less than 0.1s for all instances, while \name~uses $O(10)$s for \texttt{FM2RR} and \texttt{LPx2}, and $O(10^3)$s for \texttt{NetAcq}.
These results are expected for two reasons.
Firstly, the open-source implementation of Snowcap is overly simple, 
with no or limited support for many protocol features including longest-prefix matching, ECMP, route aggregation, etc.
Also, it restricts eBGP sessions to be established with external peers.
As a result, Snowcap finishes a round of simulation in sub-milliseconds, order-or-magnitude faster than Batfish and \name~(seconds).

Second, the synthetic migration scenarios are so simple that for most cases
Snowcap can find a safe plan without backtracking.
On the contrary, \name~targets at supporting diverse planning tasks, 
so it is necessary to explore all intermediate states with symbolic simulation.
The cost pays off when finding optimal or parallel plans but can be an overkill for sequential plans.

To validate the above hypotheses,
we modified Batfish to support incremental simulation for a limited set of configuration changes (IGP link weight change, remove or addition of BGP sessions, routing policy modifications),
and replay the search traces of Snowcap using the modified Batfish. 
As shown in \figref{fig:e2e}, using Batfish as the simulator, Snowcap's planning time significantly increases, 
and slower than \name~for most of the cases.
Results for \texttt{RRx2} and \texttt{IGPx2} are similar and shown in \appref{app:additional-results}.
This confirms our previous hypothesis why Snowcap is so fast.

For \name, \texttt{NetAcq} is substantially harder than \texttt{FM2RR} and \texttt{LPx2}.
In detail, its widespread OSPF weight modifications limit equivalence among intermediate states, 
and the number of route computed by \name~grows exponentially with the number of updates.
For instances exceeding 100 seconds, we randomly partition the updates into three groups and simulate each separately, 
and label them as \name~(grouped).
Grouping substantially reduces runtime: in some instances, even two-group partition reduces simulation time from $O(10^3)$ to $O(10)$ seconds.

\subsection{Scalability of Symbolic Simulation}
\label{sec:exp:simulation}

\begin{figure}[!t]
    \centering
    \includegraphics[width=\linewidth]{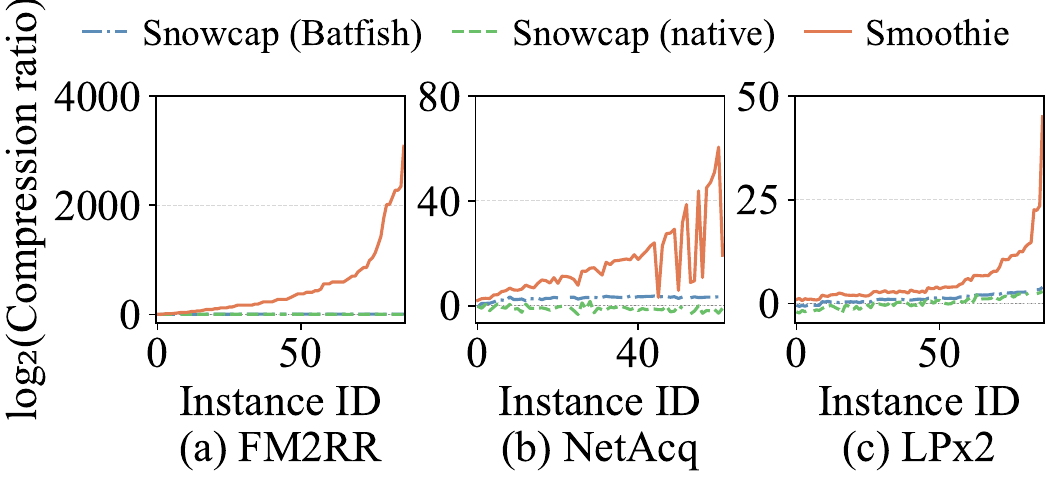}
    \vspace{-2em}
    \caption{Compression ratio to full simulation of all explored states using incremental simulation (Snowcap (native) and (Batfish)) and symbolic simulation (\name).
    }
    \vspace{-1em}
    \label{fig:simulation}
\end{figure}

This section studies how well symbolic simulation scales to a large space of intermediate states,
compared to incremental simulation as adopted in \cite{schneider2021snowcap}.

We use a metric called \emph{compression ratio}, defined as $(T_{init} \times N_{states}) / T_{total}$, where $T_{init}$ is the time to simulate the initial configuration, $N_{states}$ is the number of states explored, and $T_{total}$ is the total simulation time. 
The \emph{compression ratio} reflects how effectively a simulator can exploit equivalence in the state space.
For incremental simulation, $N_{states}$ counts the states visited during the DFS traversal and backtracking; 
while for symbolic simulation, $N_{states} = 2^{N_{updates}}$, since all intermediate states are simulated symbolically in a single execution.
When grouping is applied, $N_{states}$ is the sum of the states explored across all groups.

\figref{fig:simulation} shows the results.
The compression ratio of incremental simulation, either Snowcap (native) or (Batfish), is generally under 10.
This indicates that incremental simulation exploits only a limited number equivalent states,
and the total simulation time is proportional to the number of states explored.
In contrast, \name~achieves higher compression ratio on larger state spaces, given that the instances are sorted by increasing number of updates.
Moreover, \name~achieves higher compression ratio on \texttt{FM2RR} than on \texttt{NetAcq} and \texttt{LPx2}, since the state space of \texttt{FM2RR} is larger while having greater equivalence.
The dramatic drops of compression ratios in \figref{fig:simulation}(b) are caused by grouping.
Similar results for \texttt{RRx2} and \texttt{IGPx2} are shown in \appref{app:additional-results}.
\section{Related Work}
\label{sec_related-work}

Many prior works have studied network migrations in different contexts. We summarize them here and outline their relationship to our work.

\para{Configuration Migration} 
in the traditional networks (non-SDN) coarsely has two main branches of research.
Some works, e.g., \cite{alimi2008shadow, herrero2011network}, require routers to run both the initial and final configurations in parallel. Coexisting configurations result in coexisting data planes. This way, they gradually instruct switches to forward traffic according to the final data plane in an order that avoids forwarding loops \cite{vanbever2011seamless, vissicchio2012improving}. 
Despite being useful, the overhead and device support needed by control and data plane duplication on switches impede their deployment.

Other research \cite{clad2014computing, francois2007disruption, schneider2021snowcap, chen2023cursor, chen2025safemigration} directly finds a progressive deployment plan for configuration updates, ensuring that the network maintains correct routing and/or forwarding behaviors after each step. 
Compared to the above research, it does not require special support from routers. 
However, as mentioned earlier, unlike \name, these works are limited in their support for diverse operational intents.

\para{Physical Infrastructure Migration} 
refers to adding, removing, or swapping switches and links.
This requires maintaining bandwidth headroom to ensure the network can handle traffic demands. 
SOTA tools \cite{zhu2021network,zhao2023klotski} automate the physical transitions using different searching strategies on top of different traffic models. We'll investigate whether \name~can capture the safety constraints and diverse intents of physical infrastructure migration in the future.

\section{Discussion}

\para{Use temporary updates to expand the envelope of safe plans.}
Sometimes, no safe plan exists if we are limited to updates that results from the difference between the initial and final configurations.
Consider a variant of the example in \secref{sec_motivation-wan}:
suppose instead of receiving $DC1$'s BGP routes from routers $C$ and $D$, $DC2$ is configured with two equal-cost default static routes (i.e., \texttt{0.0.0.0/0}) towards $C$ and $D$.
Regardless of what migration plan we take, $DC2$ will always distribute its traffic towards $DC1$ to $C$ and $D$ evenly.
In such cases, we have to {\em temporarily} remove the static routes to meet the forwarding requirements.
Temporary removal of static routes is already supported by \name, but we need to add support for a broader range of temporary updates and automatically determine which ones are needed to devise a safe plan.

\para{Support for convergence behavior.}
Our symbolic simulation models the converged state of the network after applying updates, rather than transient behavior during convergence.
Complementary techniques, such as Chameleon~\cite{schneider2023taming} for BGP and~\cite{francois2007disruption} for IGP, 
decompose individual updates into smaller steps to preserve specified safety properties during convergence.
To make a plan that ensures safety both during and after re-convergence, 
we can first use \name~to order all updates, then use \cite{schneider2023taming} and \cite{francois2007disruption} to manage the safe execution of each update.
Like other control plane analysis tools~\cite{fogel2015general,beckett2017general,zhang2022symbolic}, if the network converges non-deterministically~\cite{rfc-bgp-wedgies}, \name\ will consider safety for only one of them.

\para{Use LLM for configuration migration planning.}
While planning directly from raw configurations requires an agentic system, our decoupled framework readily accommodates an LLM as a backend planner.
Preliminary experiments with ChatGPT-5.6 Sol using PDDL inputs show high first-attempt accuracy for sequential planning, 
but optimal and parallel planning often require more attempts and provide no guarantee of finding an optimal or shortest plan.
Much of this success reflects the simplicity of the synthetic cases, where the model's stated ``insert-before-remove'' strategy often suffices.
Experiments with natural-language migration descriptions and planning objectives further show that precomputed safety information improves accuracy and reduces time and token usage.
These preliminary results suggest that further research is needed to assess the practical value and reliability of LLMs for configuration migration planning.

\section{Conclusion}

\name~is a versatile configuration migration planner that supports diverse network designs and operational preferences and can also find safe parallel plans.
It achieves that by decoupling how safety constraints and preferences are encoded and how a viable plan is searched. 
Our experiments demonstrate that \name~performs well, finding sequential plans for most cases within 1 hour while compressing large migrations into no more than 3 parallel steps.

\vspace{1em}

\bibliographystyle{plain}
\bibliography{main}

@INPROCEEDINGS{schneider2021snowcap,
    isbn = {978-1-4503-8383-7},
    copyright = {In Copyright - Non-Commercial Use Permitted},
    doi = {10.3929/ethz-b-000491508},
    year = {2021-08},
    booktitle = {Proceedings of the 2021 ACM SIGCOMM Conference},
    type = {Conference Paper},
    institution = {EC},
    author = {Schneider, Tibor and Birkner, Rüdiger and Vanbever, Laurent},
    language = {en},
    address = {New York, NY},
    publisher = {Association for Computing Machinery},
    title = {Snowcap: Synthesizing Network-Wide Configuration Updates},
    PAGES = {33 - 49},
    Note = {ACM SIGCOMM 2021 Conference; Conference Location: Online; Conference Date: August 23-27, 2021}
}

@INPROCEEDINGS{schneider2023taming,
    isbn = {979-8-4007-0236-5},
    copyright = {Creative Commons Attribution 4.0 International},
    doi = {10.3929/ethz-b-000612650},
    year = {2023-09},
    booktitle = {ACM SIGCOMM '23: Proceedings of the ACM SIGCOMM 2023 Conference},
    type = {Conference Paper},
    institution = {EC and EC},
    author = {Schneider, Tibor and Schmid, Roland and Vissicchio, Stefano and Vanbever, Laurent},
    language = {en},
    address = {New York, NY},
    publisher = {Association for Computing Machinery},
    title = {Taming the transient while reconfiguring BGP},
    PAGES = {77 - 93},
    Note = {37th ACM SIGCOMM Conference (SIGCOMM 2023); Conference Location: New York, NY, USA; Conference Date: September 10-14, 2023}
}

@inproceedings{liu2018automatic,
    author = {Liu, Hongqiang Harry and Wu, Xin and Zhou, Wei and Chen, Weiguo and Wang, Tao and Xu, Hui and Zhou, Lei and Ma, Qing and Zhang, Ming},
    title = {Automatic Life Cycle Management of Network Configurations},
    year = {2018},
    isbn = {9781450359146},
    publisher = {Association for Computing Machinery},
    address = {New York, NY, USA},
    url = {https://doi.org/10.1145/3229584.3229585},
    doi = {10.1145/3229584.3229585},
    booktitle = {Proceedings of the Afternoon Workshop on Self-Driving Networks},
    pages = {29–35},
    location = {Budapest, Hungary},
    series = {SelfDN 2018}
}

@inproceedings{fogel2015general,
  title={A general approach to network configuration analysis},
  author={Fogel, Ari and Fung, Stanley and Pedrosa, Luis and Walraed-Sullivan, Meg and Govindan, Ramesh and Mahajan, Ratul and Millstein, Todd},
  booktitle={12th USENIX Symposium on Networked Systems Design and Implementation (NSDI 15)},
  pages={469--483},
  year={2015},
  publisher = {USENIX Association},
}

@inproceedings{beckett2017general,
    author = {Beckett, Ryan and Gupta, Aarti and Mahajan, Ratul and Walker, David},
    title = {A General Approach to Network Configuration Verification},
    year = {2017},
    isbn = {9781450346535},
    publisher = {Association for Computing Machinery},
    address = {New York, NY, USA},
    url = {https://doi.org/10.1145/3098822.3098834},
    doi = {10.1145/3098822.3098834},
    booktitle = {Proceedings of the Conference of the ACM Special Interest Group on Data Communication},
    pages = {155–168},
    location = {Los Angeles, CA, USA},
    series = {SIGCOMM '17}
}

@inproceedings{ye2020accuracy,
    author = {Ye, Fangdan and Yu, Da and Zhai, Ennan and Liu, Hongqiang Harry and Tian, Bingchuan and Ye, Qiaobo and Wang, Chunsheng and Wu, Xin and Guo, Tianchen and Jin, Cheng and She, Duncheng and Ma, Qing and Cheng, Biao and Xu, Hui and Zhang, Ming and Wang, Zhiliang and Fonseca, Rodrigo},
    title = {Accuracy, Scalability, Coverage: A Practical Configuration Verifier on a Global WAN},
    year = {2020},
    isbn = {9781450379557},
    publisher = {Association for Computing Machinery},
    address = {New York, NY, USA},
    url = {https://doi.org/10.1145/3387514.3406217},
    doi = {10.1145/3387514.3406217},
    booktitle = {Proceedings of the Annual Conference of the ACM Special Interest Group on Data Communication on the Applications, Technologies, Architectures, and Protocols for Computer Communication},
    pages = {599–614},
    location = {Virtual Event, USA},
    series = {SIGCOMM '20}
}

@inproceedings{zhang2022symbolic,
    author = {Zhang, Peng and Wang, Dan and Gember-Jacobson, Aaron},
    title = {Symbolic router execution},
    year = {2022},
    isbn = {9781450394208},
    publisher = {Association for Computing Machinery},
    address = {New York, NY, USA},
    url = {https://doi.org/10.1145/3544216.3544264},
    doi = {10.1145/3544216.3544264},
    booktitle = {Proceedings of the ACM SIGCOMM 2022 Conference},
    pages = {336–349},
    location = {Amsterdam, Netherlands},
    series = {SIGCOMM '22}
}

@inproceedings{wang2024expresso,
    author = {Wang, Dan and Zhang, Peng and Gember-Jacobson, Aaron},
    title = {Expresso: Comprehensively Reasoning About External Routes Using Symbolic Simulation},
    year = {2024},
    isbn = {9798400706141},
    publisher = {Association for Computing Machinery},
    address = {New York, NY, USA},
    url = {https://doi.org/10.1145/3651890.3672220},
    doi = {10.1145/3651890.3672220},
    booktitle = {Proceedings of the ACM SIGCOMM 2024 Conference},
    pages = {197–212},
    location = {Sydney, NSW, Australia},
    series = {ACM SIGCOMM '24}
}

@inproceedings{alimi2008shadow,
    author = {Alimi, Richard and Wang, Ye and Yang, Y. Richard},
    title = {Shadow configuration as a network management primitive},
    year = {2008},
    isbn = {9781605581750},
    publisher = {Association for Computing Machinery},
    address = {New York, NY, USA},
    url = {https://doi.org/10.1145/1402958.1402972},
    doi = {10.1145/1402958.1402972},
    booktitle = {Proceedings of the ACM SIGCOMM 2008 Conference on Data Communication},
    pages = {111–122},
    location = {Seattle, WA, USA},
    series = {SIGCOMM '08}
}

@book{herrero2011network,
  title={Network Mergers and Migrations: Junos Design and Implementation},
  author={Herrero, Gonzalo Gomez and Van der Ven, Jan Ant{\'o}n Bernal},
  year={2011},
  publisher={John Wiley \& Sons}
}

@inproceedings{vanbever2011seamless,
  title={Seamless network-wide IGP migrations},
  author={Vanbever, Laurent and Vissicchio, Stefano and Pelsser, Cristel and Francois, Pierre and Bonaventure, Olivier},
  booktitle={Proceedings of the ACM SIGCOMM 2011 Conference},
  pages={314--325},
  year={2011},
  publisher={Association for Computing Machinery},
  address={New York, NY, USA},
}

@article{vissicchio2012improving,
  title={Improving network agility with seamless BGP reconfigurations},
  author={Vissicchio, Stefano and Vanbever, Laurent and Pelsser, Cristel and Cittadini, Luca and Francois, Pierre and Bonaventure, Olivier},
  journal={IEEE/ACM Transactions on Networking},
  volume={21},
  number={3},
  pages={990--1002},
  year={2012},
  publisher={IEEE}
}

@article{clad2014computing,
  title={Computing minimal update sequences for graceful router-wide reconfigurations},
  author={Clad, Francois and Vissicchio, Stefano and M{\'e}rindol, Pascal and Francois, Pierre and Pansiot, Jean-Jacques},
  journal={IEEE/ACM Transactions on Networking},
  volume={23},
  number={5},
  pages={1373--1386},
  year={2014},
  publisher={IEEE}
}

@inproceedings{francois2007disruption,
  title={Disruption free topology reconfiguration in OSPF networks},
  author={Francois, Pierre and Shand, Mike and Bonaventure, Olivier},
  booktitle={IEEE INFOCOM 2007-26th IEEE International Conference on Computer Communications},
  pages={89--97},
  year={2007},
  organization={IEEE}
}

@inproceedings{xu2024relational,
    author = {Xu, Xieyang and Yuan, Yifei and Kincaid, Zachary and Krishnamurthy, Arvind and Mahajan, Ratul and Walker, David and Zhai, Ennan},
    title = {Relational Network Verification},
    year = {2024},
    isbn = {9798400706141},
    publisher = {Association for Computing Machinery},
    address = {New York, NY, USA},
    url = {https://doi.org/10.1145/3651890.3672238},
    doi = {10.1145/3651890.3672238},
    booktitle = {Proceedings of the ACM SIGCOMM 2024 Conference},
    pages = {213–227},
    location = {Sydney, NSW, Australia},
    series = {ACM SIGCOMM '24}
}

@misc{jdd,
  title = {{JDD, a pure Java BDD and Z-BDD library}},
  author = {Arash Vahidi},
  howpublished = "\url{https://bitbucket.org/vahidi/jdd/}",
}

@misc{snowcap_prototype,
  title = {{Snowcap Prototype}},
  author = {Tibor Schneider},
  howpublished = "\url{https://github.com/nsg-ethz/snowcap}",
}

@misc{batfish,
  title = {{Batfish}},
  key={batfish},
  howpublished = "\url{https://github.com/batfish/batfish}",
}

@article{singh2015jupiter,
  title={Jupiter rising: A decade of clos topologies and centralized control in google's datacenter network},
  author={Singh, Arjun and Ong, Joon and Agarwal, Amit and Anderson, Glen and Armistead, Ashby and Bannon, Roy and Boving, Seb and Desai, Gaurav and Felderman, Bob and Germano, Paulie and others},
  journal={ACM SIGCOMM computer communication review},
  volume={45},
  number={4},
  pages={183--197},
  year={2015},
  publisher={ACM New York, NY, USA}
}

@article{knight2011internet,
  title={The internet topology zoo},
  author={Knight, Simon and Nguyen, Hung X and Falkner, Nickolas and Bowden, Rhys and Roughan, Matthew},
  journal={IEEE Journal on Selected Areas in Communications},
  volume={29},
  number={9},
  pages={1765--1775},
  year={2011},
  publisher={IEEE}
}

@inproceedings{chen2023cursor,
  author={Chen, Zibin and Gao, Lixin},
  booktitle={IEEE INFOCOM 2023 - IEEE Conference on Computer Communications}, 
  title={CURSOR: Configuration Update Synthesis Using Order Rules}, 
  year={2023},
  volume={},
  number={},
  pages={1-10},
  doi={10.1109/INFOCOM53939.2023.10228930}
}

@inproceedings{chen2025safemigration,
  title={SafeMigration: Safe Large-scale Migration Planning via Symbolic Execution},
  author={Chen, Zibin and Gao, Lixin and Zhang, Ying},
  booktitle={Proceedings of the 2nd Workshop on Formal Methods Aided Network Operation},
  pages={37--42},
  year={2025}
}

@misc{rfc-route-reflection,
    series =    {Request for Comments},
    number =    4456,
    howpublished =  {RFC 4456},
    publisher = {RFC Editor},
    doi =       {10.17487/RFC4456},
    url =       {https://www.rfc-editor.org/info/rfc4456},
    author =    {Enke Chen and Tony J. Bates and Ravi Chandra},
    title =     {{BGP Route Reflection: An Alternative to Full Mesh Internal BGP (IBGP)}},
    pagetotal = 12,
    year =      2006,
    month =     apr,
}

@misc{rfc-bgp-wedgies,
    series =    {Request for Comments},
    number =    4264,
    howpublished =  {RFC 4264},
    publisher = {RFC Editor},
    doi =       {10.17487/RFC4264},
    url =       {https://www.rfc-editor.org/info/rfc4264},
    author =    {Tim A. Griffin and Geoff Huston},
    title =     {{BGP Wedgies}},
    pagetotal = 10,
    year =      2005,
    month =     nov,
}

@inproceedings{li2024general,
    author = {Li, Ruihan and Yuan, Yifei and Ye, Fangdan and Liu, Mengqi and Yang, Ruizhen and Yu, Yang and Guo, Tianchen and Ma, Qing and Zeng, Xianlong and Xu, Chenren and Cai, Dennis and Zhai, Ennan},
    title = {A General and Efficient Approach to Verifying Traffic Load Properties under Arbitrary k Failures},
    year = {2024},
    isbn = {9798400706141},
    publisher = {Association for Computing Machinery},
    address = {New York, NY, USA},
    url = {https://doi.org/10.1145/3651890.3672246},
    doi = {10.1145/3651890.3672246},
    booktitle = {Proceedings of the ACM SIGCOMM 2024 Conference},
    pages = {228–243},
    numpages = {16},
    location = {Sydney, NSW, Australia},
    series = {ACM SIGCOMM '24}
}

@article{aeronautiques1998pddl,
  title={Pddl—the planning domain definition language},
  author={Aeronautiques, Constructions and Howe, Adele and Knoblock, Craig and McDermott, ISI Drew and Ram, Ashwin and Veloso, Manuela and Weld, Daniel and Sri, David Wilkins and Barrett, Anthony and Christianson, Dave and others},
  journal={Technical Report, Tech. Rep.},
  year={1998}
}

@inproceedings{kautz1992planning,
  title={Planning as Satisfiability.},
  author={Kautz, Henry A and Selman, Bart and others},
  booktitle={ECAI},
  volume={92},
  pages={359--363},
  year={1992}
}

@inproceedings{gerevini2002lpg,
  title={LPG: A Planner Based on Local Search for Planning Graphs with Action Costs.},
  author={Gerevini, Alfonso and Serina, Ivan and others},
  booktitle={Aips},
  volume={2},
  pages={281--290},
  year={2002}
}

@article{helmert2006fast,
  title={The fast downward planning system},
  author={Helmert, Malte},
  journal={Journal of Artificial Intelligence Research},
  volume={26},
  pages={191--246},
  year={2006}
}

@article{edelkamp2001mips,
  title={MIPS: The model-checking integrated planning system},
  author={Edelkamp, Stefan and Helmert, Malte},
  journal={AI magazine},
  volume={22},
  number={3},
  pages={67--67},
  year={2001}
}

@inproceedings{torralba2014symba,
  title={SymBA*: A symbolic bidirectional A* planner},
  author={Torralba, Alvaro and Alc{\'a}zar, Vidal and Borrajo, Daniel and Kissmann, Peter and Edelkamp, Stefan},
  booktitle={International Planning Competition},
  pages={105--108},
  year={2014}
}

@inproceedings{speck2020symk,
  title = {Symbolic Top-k Planning},
  author = {Speck, David and Mattm{\"u}ller, Robert and Nebel, Bernhard},
  editor = {Conitzer, Vincent and Sha, Fei},
  booktitle = {Proceedings of the Thirty-Fourth AAAI Conference on Artificial Intelligence (AAAI 2020)},
  publisher = {AAAI Press},
  year = {2020},
  pages = {9967--9974}
}

@article{rintanen2014madagascar,
  title={Madagascar: Scalable planning with SAT},
  author={Rintanen, Jussi},
  journal={Proceedings of the 8th International Planning Competition (IPC-2014)},
  volume={21},
  pages={1--5},
  year={2014}
}

@inproceedings{holler2022encoding,
  title={Encoding lifted classical planning in propositional logic},
  author={H{\"o}ller, Daniel and Behnke, Gregor},
  booktitle={Proceedings of the International Conference on Automated Planning and Scheduling},
  volume={32},
  pages={134--144},
  year={2022}
}

@article{cashmore2020planning,
  title={Planning for hybrid systems via satisfiability modulo theories},
  author={Cashmore, Michael and Magazzeni, Daniele and Zehtabi, Parisa},
  journal={Journal of Artificial Intelligence Research},
  volume={67},
  pages={235--283},
  year={2020}
}

@misc{fast-downward-website,
  author = {Helmert, Malte},
  title = {{Fast Downward}},
  howpublished={\url{https://www.fast-downward.org/latest/}},
}

@misc{enhsp-website,
  author = {Enrico Scala},
  title = {{ENHSP: Expressive Numeric Heuristic Search Planner}},
  howpublished={\url{https://sites.google.com/view/enhsp/}},
}

@book{haslum2019pddl,
  title={An introduction to the planning domain definition language},
  author={Haslum, Patrik and Lipovetzky, Nir and Magazzeni, Daniele and Muise, Christian and Brachman, Ronald and Rossi, Francesca and Stone, Peter},
  volume={13},
  year={2019},
  publisher={Springer}
}

@misc{metis,
  title={METIS},
  key={metis},
  howpublished={\url{http://glaros.dtc.umn.edu/gkhome/views/metis}}
}

@inproceedings{lin2025centralium,
  title={Centralium: A hybrid route-planning framework for large-scale data center network migrations},
  author={Lin, Yikai and Gawish, Mohab and Tseng, Shih-Hao and Gao, Lixin and Zhao, Cen and Tracey, John and Shao, Sunyi and Kim, Hyojeong and Zhang, Ying},
  booktitle={Proceedings of the ACM SIGCOMM 2025 Conference},
  pages={395--408},
  year={2025}
}

@inproceedings{zhu2021network,
    author = {Zhu, Hang and Gupta, Varun and Ahuja, Satyajeet Singh and Tian, Yuandong and Zhang, Ying and Jin, Xin},
    title = {Network planning with deep reinforcement learning},
    year = {2021},
    isbn = {9781450383837},
    publisher = {Association for Computing Machinery},
    address = {New York, NY, USA},
    url = {https://doi.org/10.1145/3452296.3472902},
    doi = {10.1145/3452296.3472902},
    booktitle = {Proceedings of the 2021 ACM SIGCOMM 2021 Conference},
    pages = {258–271},
    numpages = {14},
    location = {Virtual Event, USA},
    series = {SIGCOMM '21}
}

@inproceedings{zhao2023klotski,
    author = {Zhao, Yihao and Zhang, Xiaoxiang and Zhu, Hang and Zhang, Ying and Wang, Zhaodong and Tian, Yuandong and Nikulkov, Alex and Ferreira, Joao and Liu, Xuanzhe and Jin, Xin},
    title = {Klotski: Efficient and Safe Network Migration of Large Production Datacenters},
    year = {2023},
    isbn = {9798400702365},
    publisher = {Association for Computing Machinery},
    address = {New York, NY, USA},
    url = {https://doi.org/10.1145/3603269.3604818},
    doi = {10.1145/3603269.3604818},
    booktitle = {Proceedings of the ACM SIGCOMM 2023 Conference},
    pages = {783–797},
    numpages = {15},
    location = {New York, NY, USA},
    series = {ACM SIGCOMM '23}
}

@inproceedings{yuan2025new,
    author = {Yuan, Yifei and Ye, Fangdan and Li, Yifan and Zhang, Jingkai and Liu, Mengqi and Sang, Yuyang and Yang, Ruizhen and She, Duncheng and Ye, Zhiqing and Guo, Tianchen and Zhu, Xiaobo and Tang, Xinji and Jia, Li and Guan, Zhongyu and Su, Lingpeng and Wang, Ci and Feng, Ruiyang and Wu, Shuo and Xie, Zhonghui and Jin, Cheng and Zhang, Peng and Ma, Qing and Zeng, Xianlong and Cai, Dennis and Zhai, Ennan},
    title = {New Evolution of Hoyan: Enhancing Scalability, Usability, and Accuracy for Alibaba's Global WAN Verification},
    year = {2025},
    isbn = {9798400715242},
    publisher = {Association for Computing Machinery},
    address = {New York, NY, USA},
    url = {https://doi.org/10.1145/3718958.3754343},
    doi = {10.1145/3718958.3754343},
    booktitle = {Proceedings of the ACM SIGCOMM 2025 Conference},
    pages = {809–825},
    numpages = {17},
    location = {S\~{a}o Francisco Convent, Coimbra, Portugal},
    series = {SIGCOMM '25}
}

@article{rintanen2006planning,
  title={Planning as satisfiability: parallel plans and algorithms for plan search},
  author={Rintanen, Jussi and Heljanko, Keijo and Niemel{\"a}, Ilkka},
  journal={Artificial Intelligence},
  volume={170},
  number={12-13},
  pages={1031--1080},
  year={2006},
  publisher={Elsevier}
}

@article{rintanen2012planning,
  title={Planning as satisfiability: Heuristics},
  author={Rintanen, Jussi},
  journal={Artificial intelligence},
  volume={193},
  pages={45--86},
  year={2012},
  publisher={Elsevier}
}

@article{hart1968formal,
  title={A formal basis for the heuristic determination of minimum cost paths},
  author={Hart, Peter E and Nilsson, Nils J and Raphael, Bertram},
  journal={IEEE transactions on Systems Science and Cybernetics},
  volume={4},
  number={2},
  pages={100--107},
  year={1968},
  publisher={IEEE}
}

@misc{z3,
  author = {{Microsoft Research}},
  title = {{Z3 Theorem Prover}},
  howpublished = {\url{https://github.com/Z3Prover/z3}},
}

@article{blum1997fast,
  title={Fast planning through planning graph analysis},
  author={Blum, Avrim L and Furst, Merrick L},
  journal={Artificial intelligence},
  volume={90},
  number={1-2},
  pages={281--300},
  year={1997},
  publisher={Elsevier}
}

@inproceedings{coles2010forward,
  title={Forward-chaining partial-order planning},
  author={Coles, Amanda and Coles, Andrew and Fox, Maria and Long, Derek},
  booktitle={Proceedings of the International Conference on Automated Planning and Scheduling},
  volume={20},
  pages={42--49},
  year={2010}
}

@inproceedings{shaik2023optimal,
  title={Optimal layout synthesis for quantum circuits as classical planning},
  author={Shaik, Irfansha and Van De Pol, Jaco},
  booktitle={2023 IEEE/ACM International Conference on Computer Aided Design (ICCAD)},
  pages={1--9},
  year={2023},
  organization={IEEE}
}

@inproceedings{vallati2016efficient,
  title={Efficient macroscopic urban traffic models for reducing congestion: A PDDL+ planning approach},
  author={Vallati, Mauro and Magazzeni, Daniele and De Schutter, Bart and Chrpa, Luk{\'a}s and McCluskey, Thomas},
  booktitle={Proceedings of the AAAI conference on artificial intelligence},
  volume={30},
  number={1},
  year={2016}
}

@inproceedings{mccluskey2017embedding,
  title={Embedding automated planning within urban traffic management operations},
  author={McCluskey, Thomas and Vallati, Mauro},
  booktitle={Proceedings of the International Conference on Automated Planning and Scheduling},
  volume={27},
  pages={391--399},
  year={2017}
}

@inproceedings{ramanathan2023practical,
  title={Practical intent-driven routing configuration synthesis},
  author={Ramanathan, Sivaramakrishnan and Zhang, Ying and Gawish, Mohab and Mundada, Yogesh and Wang, Zhaodong and Yun, Sangki and Lippert, Eric and Taha, Walid and Yu, Minlan and Mirkovic, Jelena},
  booktitle={20th USENIX Symposium on Networked Systems Design and Implementation (NSDI 23)},
  pages={629--644},
  year={2023}
}

@inproceedings{valentini2020temporal,
  title={Temporal planning with intermediate conditions and effects},
  author={Valentini, Alessandro and Micheli, Andrea and Cimatti, Alessandro},
  booktitle={Proceedings of the AAAI Conference on Artificial Intelligence},
  volume={34},
  number={06},
  pages={9975--9982},
  year={2020}
}

@inproceedings{bit2023experimenting,
  title={Experimenting with Lifted Plan-Space Planning as Scheduling: Aries in the 2023 IPC},
  author={Bit-Monnot, Arthur},
  booktitle={2023 International Planning Competition at the 33rd International Conference on Automated Planning and Scheduling},
  year={2023}
}

@inproceedings{kautz1996pushing,
  title={Pushing the envelope: Planning, propositional logic, and stochastic search},
  author={Kautz, Henry and Selman, Bart},
  booktitle={Proceedings of the national conference on artificial intelligence},
  pages={1194--1201},
  year={1996}
}

@inproceedings{wright2018compiling,
  title={Compiling Away Soft Trajectory Constraints in Planning.},
  author={Wright, Benedict and Mattm{\"u}ller, Robert and Nebel, Bernhard},
  booktitle={KR},
  pages={474--483},
  year={2018}
}

@inproceedings{percassi2019compiling,
  title={On compiling away PDDL3 soft trajectory constraints without using automata},
  author={Percassi, Francesco and Gerevini, Alfonso E},
  booktitle={Proceedings of the International Conference on Automated Planning and Scheduling},
  volume={29},
  pages={320--328},
  year={2019}
}

@inproceedings{abdulaziz2023formally,
  title={Formally verified SAT-based AI planning},
  author={Abdulaziz, Mohammad and Kurz, Friedrich},
  booktitle={Proceedings of the AAAI Conference on Artificial Intelligence},
  volume={37},
  number={12},
  pages={14665--14673},
  year={2023}
}

@article{richter2010lama,
  title={The LAMA planner: Guiding cost-based anytime planning with landmarks},
  author={Richter, Silvia and Westphal, Matthias},
  journal={Journal of Artificial Intelligence Research},
  volume={39},
  pages={127--177},
  year={2010}
}

@article{hansen2007anytime,
  title={Anytime heuristic search},
  author={Hansen, Eric A and Zhou, Rong},
  journal={Journal of Artificial Intelligence Research},
  volume={28},
  pages={267--297},
  year={2007}
}

@inproceedings{helmert2009landmarks,
  title={Landmarks, critical paths and abstractions: what's the difference anyway?},
  author={Helmert, Malte and Domshlak, Carmel},
  booktitle={Proceedings of the international conference on automated planning and scheduling},
  volume={19},
  pages={162--169},
  year={2009}
}

\appendix
\section{Additional Evaluation Results}
\label{app:additional-results}

\begin{figure}[ht]
    \centering
    \includegraphics[width=\linewidth]{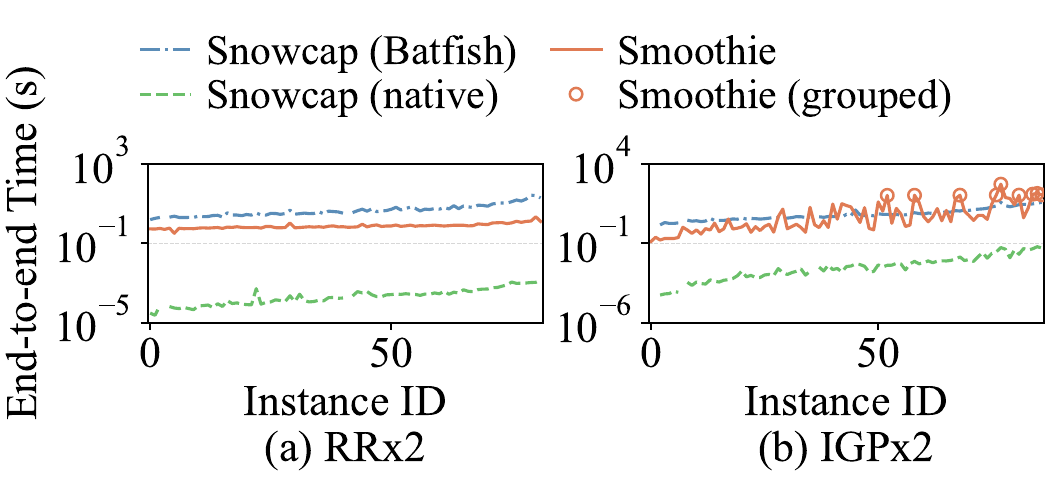}
    \caption{End-to-end time to find any safe sequential plan for \texttt{RRx2} and \texttt{IGPx2} instances using Snowcap, Batfish, and \name. For instances where \name~ does not finish in 100 seconds, we randomly divide all updates into 3 groups and separately simulate each group. Such instances are tagged as \name~(grouped).}
    \label{fig:e2e-app}
\end{figure}

\begin{figure}[ht]
    \centering
    \includegraphics[width=\linewidth]{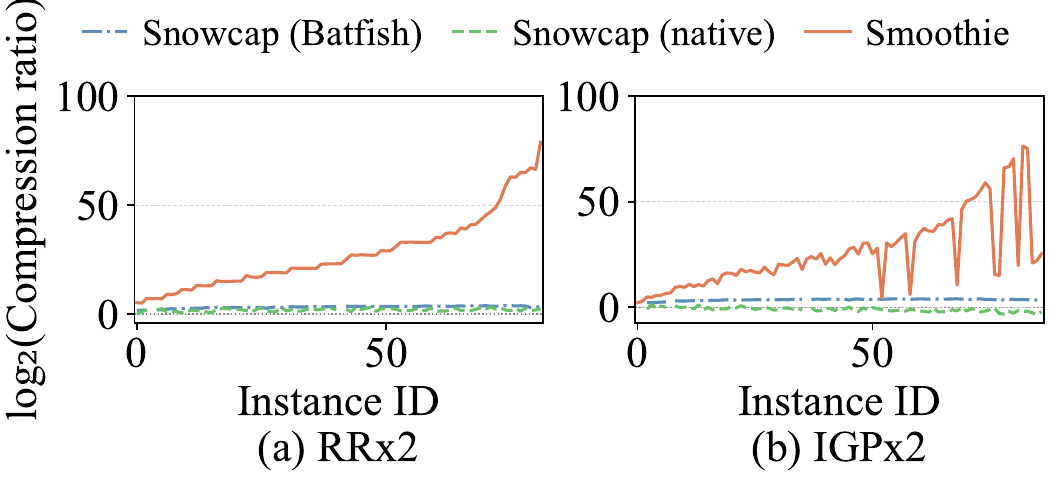}
    \caption{Compression ratio ($T_{init} \times N_{states} / T_{total}$) of incremental simulation (Snowcap and Batfish) and symbolic simulation (\name) for \texttt{RRx2} and \texttt{IGPx2} instances, compared to full simulation of all considered intermediate states.}
    \label{fig:simulation-app}
\end{figure}

\figref{fig:e2e-app} and \figref{fig:simulation-app} show the additional results on \texttt{RRx2} and \texttt{IGPx2} for evaluations introduced in \secref{sec:exp:comparison} and \secref{sec:exp:simulation}, respectively.

\end{document}